\documentclass[twocolumn]{aastex701}
\usepackage{amsmath, amssymb, graphicx, xcolor}
\usepackage{hyperref}

\newcommand\T{$T_{\rm eff}$}

\shorttitle{A test of FeH line parameters using DR19 APOGEE spectra of benchmark M dwarfs}
\shortauthors{Silva-Andrade et al.}
\usepackage{xcolor} 

\definecolor{darkgreen}{rgb}{0.0, 0.5, 0.0} 

\begin{document}

\title{A Test of FeH Line Parameters using DR19 APOGEE spectra of Benchmark M Dwarfs}

\author[0009-0000-2228-8234]{Anderson Silva-Andrade}
\affiliation{Departamento de F\'isica, Universidade Federal de Sergipe, Av. Marcelo Deda Chagas, S/N, 49107-230, S\~ao Crist\'ov\~ao, SE, Brazil}
\affiliation{Astrophysics Research Institute, Liverpool John Moores University, 146 Brownlow Hill, Liverpool, Merseyside L3 5RF, UK}
\email[show]{anderson.silva1@academico.ufs.br}

\author[0000-0002-7883-5425]{Diogo Souto}
\affiliation{Departamento de F\'isica, Universidade Federal de Sergipe, Av. Marcelo Deda Chagas, S/N, 49107-230, S\~ao Crist\'ov\~ao, SE, Brazil}
\email{diogosouto@academico.ufs.br}

\author[0000-0001-6476-0576]{Katia Cunha}
\affiliation{Steward Observatory, University of Arizona, 933 North Cherry Avenue, Tucson, AZ 85721-0065, USA}
\affiliation{Observatório Nacional/MCTIC, R. Gen. José Cristino, 77,  20921-400, Rio de Janeiro, Brazil}
\email{YYY}

\author[0000-0002-0134-2024]{Verne V. Smith}
\affiliation{NSF’s NOIRLab, 950 N. Cherry Ave. Tucson, AZ 85719 USA}
\email{ZZZ}

\author[0000-0002-0084-572X]{Carlos Allende Prieto}
\affiliation{Instituto de Astrofísica de Canarias, C/ Vía Láctea, s/n, E-38205 San Cristóbal de La Laguna, Santa Cruz de Tenerife, Spain}
\email{TTT}

\author[0000-0002-2244-0897]{Ricardo P. Schiavon}
\affiliation{Astrophysics Research Institute, Liverpool John Moores University, 146 Brownlow Hill, Liverpool, Merseyside L3 5RF, UK}
\email{TTT}

\author[0000-0003-0506-8269]{Verónica Loaiza-Tacuri}
\affiliation{Departamento de F\'isica, Universidade Federal de Sergipe, Av. Marcelo Deda Chagas, S/N, 49107-230, S\~ao Crist\'ov\~ao, SE, Brazil}
\email{TTT}

\author[0000-0003-0012-9093]{Aida Behmard}
\affiliation{Center for Computational Astrophysics, Flatiron Institute, 162 Fifth Ave, New York, NY 10010, USA}
\affiliation{American Museum of Natural History, 200 Central Park West, Manhattan, NY 10024, USA}
\email{TTT}

\author[0000-0002-3601-133X]{Dmitry Bizyaev} 
\affiliation{Apache Point Observatory and New Mexico State University, Sunspot, NM 88349, USA}
\email{TTT}

\author[0000-0003-3410-5794]{Ilija Medan}
\affiliation{Department of Physics and Astronomy, Vanderbilt University, Nashville, TN 37235, USA}
\email{TTT}

\author[0000-0002-8280-4808]{Dan Qiu}
\affiliation{Kavli Institute for Astronomy and Astrophysics, Peking University, Beĳing 100871, People’s Republic of China}
\email{TTT}

\author[0000-0002-0149-1302]{Bárbara Rojas-Ayala}
\affiliation{Instituto de Alta Investigación, Universidad de Tarapacá, Casilla 7D, Arica, Chile}
\email{TTT}

\begin{abstract}

Recent studies have suggested a mismatch of up to 0.20 dex between iron abundances derived from Fe I and FeH lines in the H-band spectra of M dwarfs, and in this work we investigate the nature of this possible offset.
We analyze near-infrared H-band APOGEE spectra of stars in 18 binaries composed of a G-dwarf primary and an M-dwarf secondary, together with four M-dwarf stars having measured angular diameters from the literature, and six M-dwarf members of the Hyades and Coma Berenices open clusters. These three families of benchmarks were used to evaluate the FeH line list and check for possible systematic uncertainties in the FeH $gf$-values.
Our tests used 1-D LTE plane-parallel model atmospheres, a radiative transfer code, and the baseline APOGEE spectral line list to derive metallicities for the binary G-dwarf primaries using Fe I lines, while stellar parameters and metallicities for the M dwarfs used both FeH and Fe I lines. 
The mean metallicity obtained for the Hyades M-dwarfs was $\langle$[Fe/H]$\rangle$=+0.08$\pm$0.04, and for Coma Berenices $\langle$[Fe/H]$\rangle$=+0.02$\pm$0.08. 
The metallicities of the G- and M-dwarfs in binary systems showed excellent agreement (0.06 dex), and the mean metallicities for the open clusters were also consistent with literature values from high-resolution optical analyses.
We investigated the consistency between the spectroscopic and interferometric \T~scales, finding agreement within the uncertainties.
Forcing full agreement between the spectroscopic and interferometric \T~scales resulted in a poorer match for the metallicities in the binaries and the open clusters. We conclude that the best overall concordance is obtained when no adjustments are made to the FeH $gf$-values, which are based on the Hargreaves et al. (2010) line list.
\end{abstract}


\section{Introduction}\label{sec:intro}

The near-infrared (NIR) spectra of the cool M-dwarf stars exhibit fewer and weaker molecular bands when compared to the optical regime \citep{allard2000} and are particularly well-suited for deriving stellar parameters and abundances.  Although molecular absorption is generally diminished, significant absorption from molecules such as H$_2$O and FeH persists in the NIR and can be used as tools with which to derive precise effective temperatures, surface gravities, and metallicities for M-dwarf stars.  

Iron hydride produces significant absorption bands in the spectra of M-dwarf stars and becomes particularly prominent in the NIR regions of the spectra. It was first noted as an unidentified absorption feature by  \cite{Wing1969} in the M-dwarf Wolf 359 (M6 V) at a wavelength of $\lambda\sim$0.99$\mu$m and subsequently identified as FeH by \cite{Nordh1977} through spectra of M dwarfs, as well as S stars, compared to laboratory spectra of FeH; this feature is often referred to as the Wing-Ford band. \cite{Phillips1988} analyzed the FeH spectrum in detail and conclusively identified the Wing-Ford band as arising from the electronic transition F$^{4}\Delta$ – X$^{4}\Delta$. In M dwarfs, such FeH lines have been shown to be sensitive to activity and used to derive mean magnetic fields (\citealt{Reiners2006}, \citealt{Shulyak2019}). In addition to the Wing-Ford band, FeH also exhibits prominent absorption bands at $\lambda\sim$1.35$\mu$m, arising from the $E^{4}\Pi$--$X^{4}\Delta$ transition (\citealt{Cushing2003,Balfour2004}), and at $\lambda\sim$1.6$\mu$m, corresponding to the $E^{4}\Pi$--$A^{4}\Pi$ transition (\citealt{Wallace2000,Balfour2004})

The use of FeH in quantitative spectroscopic analyses of M dwarfs has become more common recently due to increasing access to NIR high-resolution spectra (e.g., \citealt{Lindgren2016} and \citealt{LindgrenHeiter2017}).
In particular, the Apache Point Observatory Galactic Evolution Experiment (APOGEE: \citealt{Majewski2017}) survey has resulted in the use of FeH as a quantitative spectroscopic diagnostic tool for much larger M-dwarf samples. Within the APOGEE H-band spectral window ($\lambda1.51$--$1.70,\mu$m), \citet{Souto2017} and \citet{Souto2020} demonstrated that FeH features can be used to infer effective temperature, surface gravity, and metallicity. However, \citet{Souto2017} and \citet{Souto2020} also report a systematic difference between iron abundances derived from Fe I and FeH lines, with FeH-based abundances being on average 0.10--0.15 dex lower than those inferred from atomic Fe I transitions. This discrepancy highlights the need for further investigation into the consistency of Fe I- and FeH-based abundance determinations, as well as the molecular data used to model FeH transitions.

This work aims to determine whether the mismatch between iron abundances derived from Fe I and FeH lines reported in the literature is associated with uncertainties in the adopted FeH line lists, systematic effects in the spectroscopic analysis, or spurious abundance offsets. 
To address this question, we selected a sample of benchmark M dwarfs in binary systems, together with M dwarfs with measured angular diameters and members of open clusters. These objects provide critical benchmarks for testing and validating atomic and molecular line lists, spectroscopic methodologies, and stellar parameters for cool stars.

Binary systems are fundamental tools in stellar astrophysics.
Such systems are relatively common: approximately half (50\% $\pm$ 4\%) of solar-type (F and G) stars in the solar neighborhood reside in binary systems \citep{Duquennoy_1991,Raghavan_2010}. Among low-mass stars such as M dwarfs, the fraction found in multiple systems is around 26.8\% $\pm$ 1.4\% \citep{Winters_2019}. With orbital periods spanning roughly $10$ to $10^8$ years \citep{El_Badry} and separations ranging from approximately one solar radius (5$\times$10$^{-3}$ AU) to $10^4$ AU \citep{Zachary_2020}, binary systems are highly valuable as benchmark stars. 
This is primarily because the stars in a binary system form simultaneously from the same gas and dust cloud, implying that they share nearly identical ages and initial chemical compositions.
Within wide binary systems containing M dwarfs, FG-type primaries are particularly valuable as reference standards for validating the metallicity of their cooler companions. 

Several studies have used the relationship between M-dwarf secondaries and FGK-type primaries to define metallicity scales using near-infrared spectroscopy through different techniques. For example, \citet{RojasAyala2012} employed equivalent widths (EWs) derived from low-resolution (R$\sim$2,000) K-band spectra to estimate metallicities for 18 M dwarfs. Subsequent investigations, including those by \citet{Newton2014}, \citet{Muirhead2014}, \citet{Terrien2015}, and \citet{Mann2013}, adopted similar methodologies while exploring different near-infrared bands and spectral lines. 

In addition to binary stars, M dwarfs that are members of open clusters are valuable benchmarks for evaluating metallicity scales and to probe chemical homogeneity in the gas that formed the stars, versus internal errors in the abundance determinations. If the sample spans a range of spectral types, trends in chemical abundances with effective temperature can reveal potential systematics in the determinations (\citealt{Wanderley2023}, \citealt{Souto2021}, \citealt{Grilo2024}, \citealt{Vilar2025}). Finally, M dwarfs with measured angular diameters can serve as benchmarks in defining effective temperature scales \citep{Souto2020}.

This paper adopts the methodology of these previous works to analyze APOGEE DR19 spectra in a sample of benchmark M dwarf stars. The sample consists of M dwarfs in binary systems having a G-type primary, M dwarfs with measured angular diameters in the literature, and M dwarf members of the Hyades and Coma Berenices open clusters, with the focus being a detailed investigation and assessment of possible offsets in the transition probabilities ($gf$-values) of the FeH lines in the analysis of APOGEE spectra and their impact, if any, on the derivation of fundamental stellar parameters for M-dwarf stars. 

Section \ref{sec:FeH} discusses the FeH line lists adopted in our analysis. In Section \ref{sec:obsdata}, we describe the observational data. 
Section \ref{sec:abundana} introduces abundance analysis and the derivation of the stellar parameters. Section \ref{sec:baseline} discusses our results and describes the validation of the FeH line list using benchmark stars. Finally, in Section \ref{sec:conclusions}, we summarize our conclusions.

\section{The FeH Line Lists in the Near-Infrared}
\label{sec:FeH}

The spectra of M dwarfs show contributions from transitions of iron hydride molecules which arise from two quartet systems (\citealt{Crozet2023}), F$^{4}\Delta$ – X$^{4}\Delta$, corresponding to the Wing-Ford band at $\lambda\sim$0.99$\mu$m \citep[e.g.,][]{Schiavon1997,Wende2010}, the E$^{4}\Pi$ – X$^{4}\Delta$ bands at $\sim$1.35$\mu$m (\citealt{Cushing2003,Balfour2004}), and the E$^{4}\Pi$ – A$^{4}\Pi$ bands at $\lambda\sim$1.6$\mu$m \citep{Wallace2000,Balfour2004}.

In this study, the line list for the F$^{4}\Delta$ – X$^{4}\Delta$ (hereafter F-X$^{4}\Delta$) electronic transition was taken from the theoretical work of \cite{Dulick2003}, who calculated FeH line intensities and Einstein A-values from ab initio calculations of electronic transition dipole moment. In that study, the FeH line list was used to compute molecular opacities for M- and L-dwarf atmospheres. \cite{Wende2010} used high-resolution spectra of the M dwarf G1002 to empirically confirm the wavelengths of the FeH lines from the \cite{Dulick2003} in the range between  $\lambda$0.9898 $\mu$m to 1.0766 $\mu$m, providing corrections to the wavelengths of the FeH transitions. Both line lists from \cite{Dulick2003} and \cite{Wende2010} are noted in the MoLLIST page maintained by \cite{Bernarth2020} at \url{https://uwaterloo.ca/bernath-lab/mollist-molecular-line-lists-intensities-and-spectra}.

For the FeH E$^{4}\Pi$ – A$^{4}\Pi$ (hereafter E-A$^{4}\Pi$) transition, we used \cite{Hargreaves2010}. This line list is based on an FeH laboratory spectrum from the FTS of the McMath-Pierce solar Telescope (see \citealt{Phillips1988} for details), and ab initio calculations for band strength (similarly to what was previously done by \citealt{Dulick2003}). In addition, \cite{Hargreaves2010} analyzed high-resolution spectra of M and L dwarfs obtained with the Phoenix spectrograph on Gemini South. 

\cite{Hargreaves2010}'s line list covers the range between $\lambda$1.58 to 2 $\micron$ and contains 292 assigned transitions from \cite{Balfour2004} and 6065 unassigned transitions, the latter with an estimated lower state energy of 2250~cm$^{-1}$ (0.279~eV). They provide Einstein A-values for the classified FeH transitions that were then used to compute $gf$-values in \cite{Souto2017}. For the unclassified transitions, \cite{Souto2017} used the line intensities presented in \cite{Hargreaves2010} to derive Einstein A-values as described in \cite{Simecova2006} and subsequent $gf$-values. Explicit uncertainty estimates for the $gf$-values calculated in this way using the results from \cite{Hargreaves2010} are not given, and one goal of this study is to use a benchmark M-dwarf sample to ascertain possible empirical offsets to the FeH $gf$-values used.

\begin{figure*}[!t]
    \centering
    \includegraphics[width=0.45\textwidth]{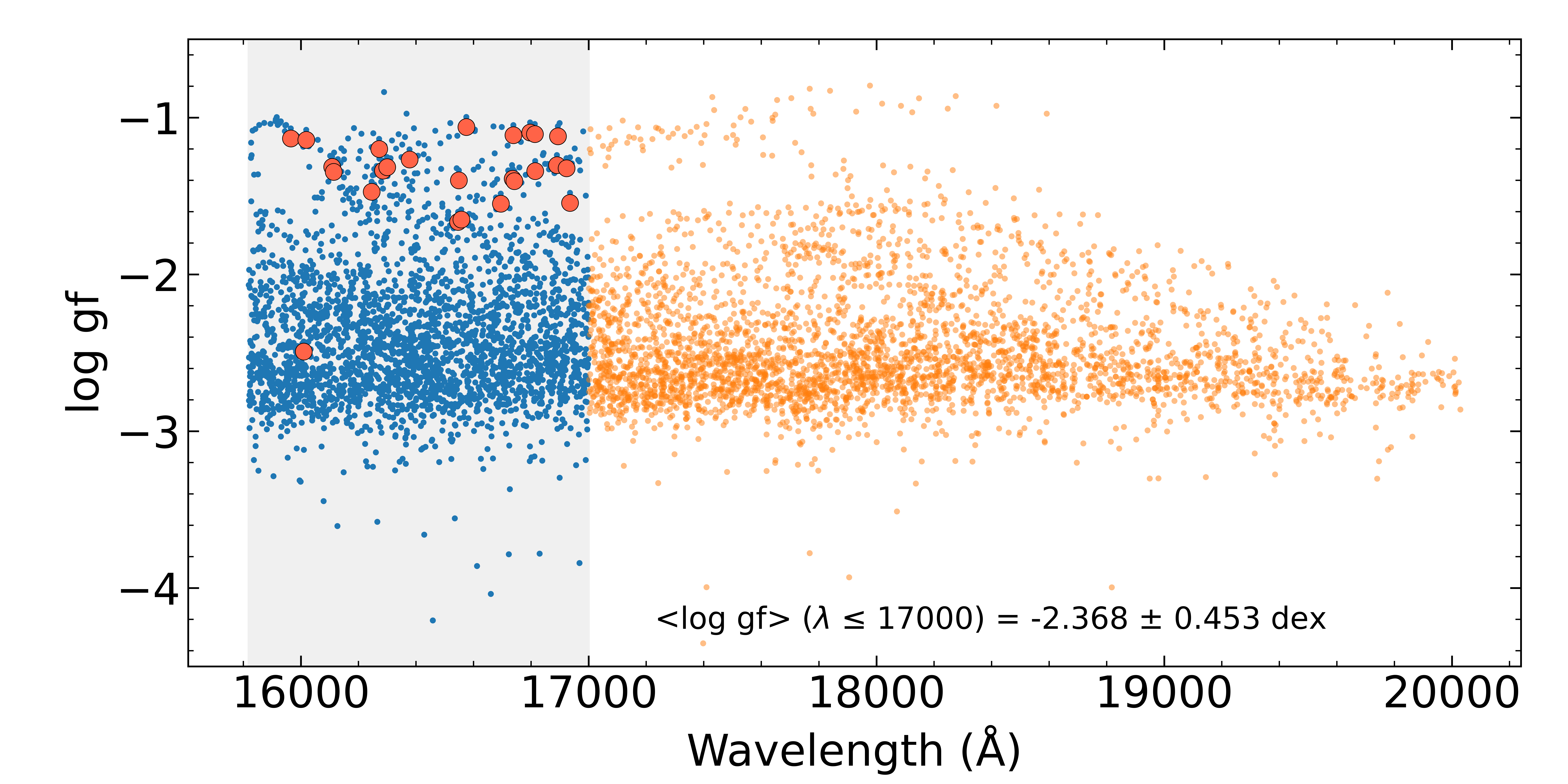}
    \includegraphics[width=0.45\textwidth]{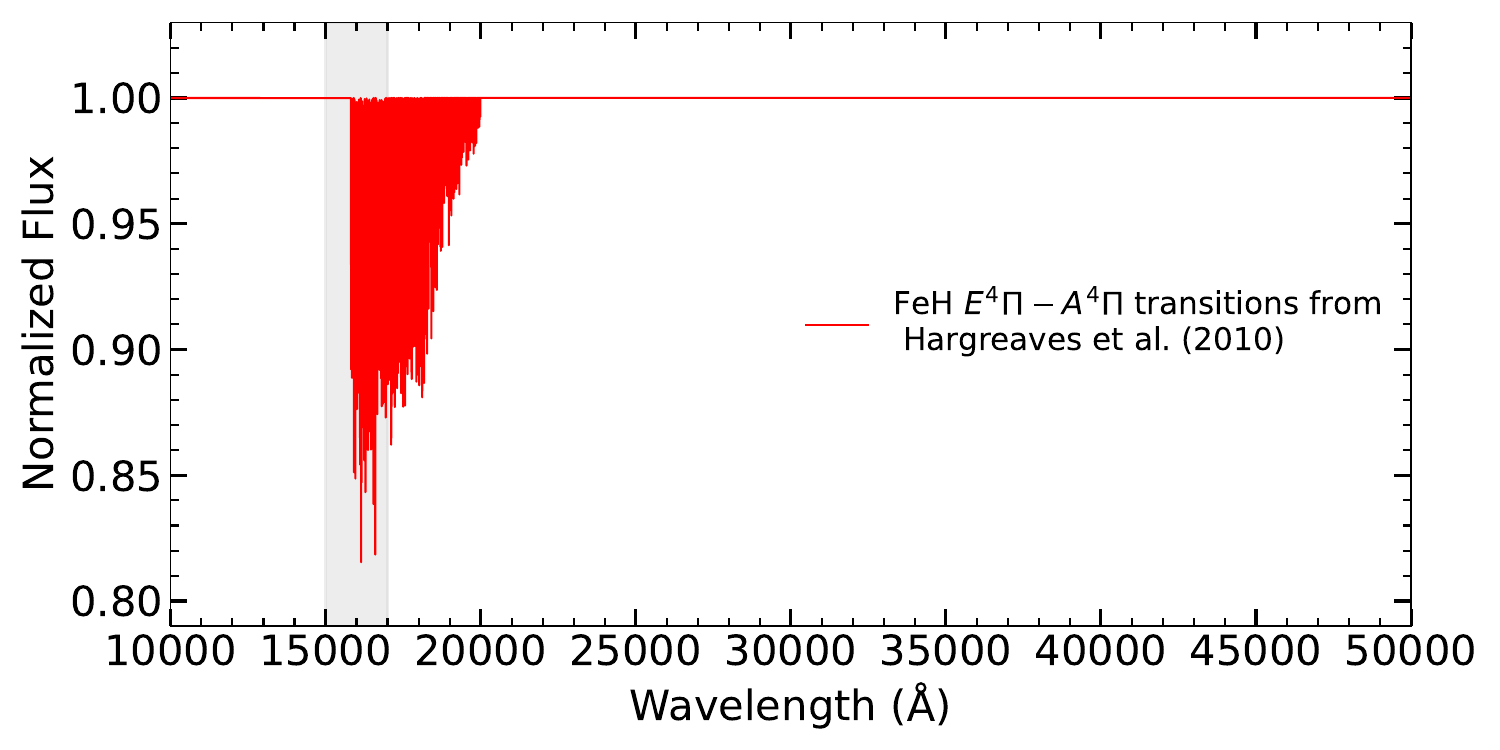}    
    \includegraphics[width=0.45\textwidth]{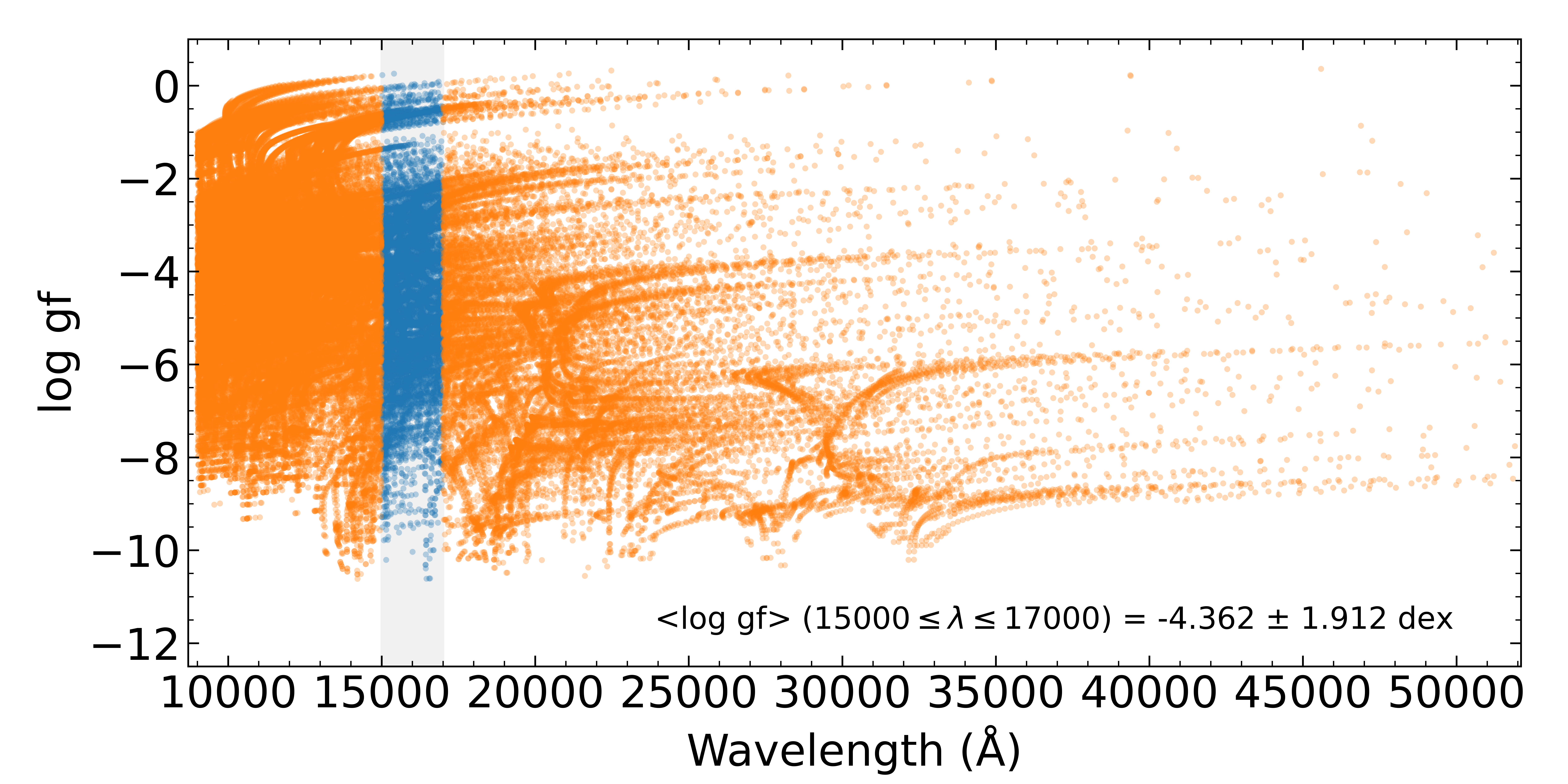}    
    \includegraphics[width=0.45\textwidth]{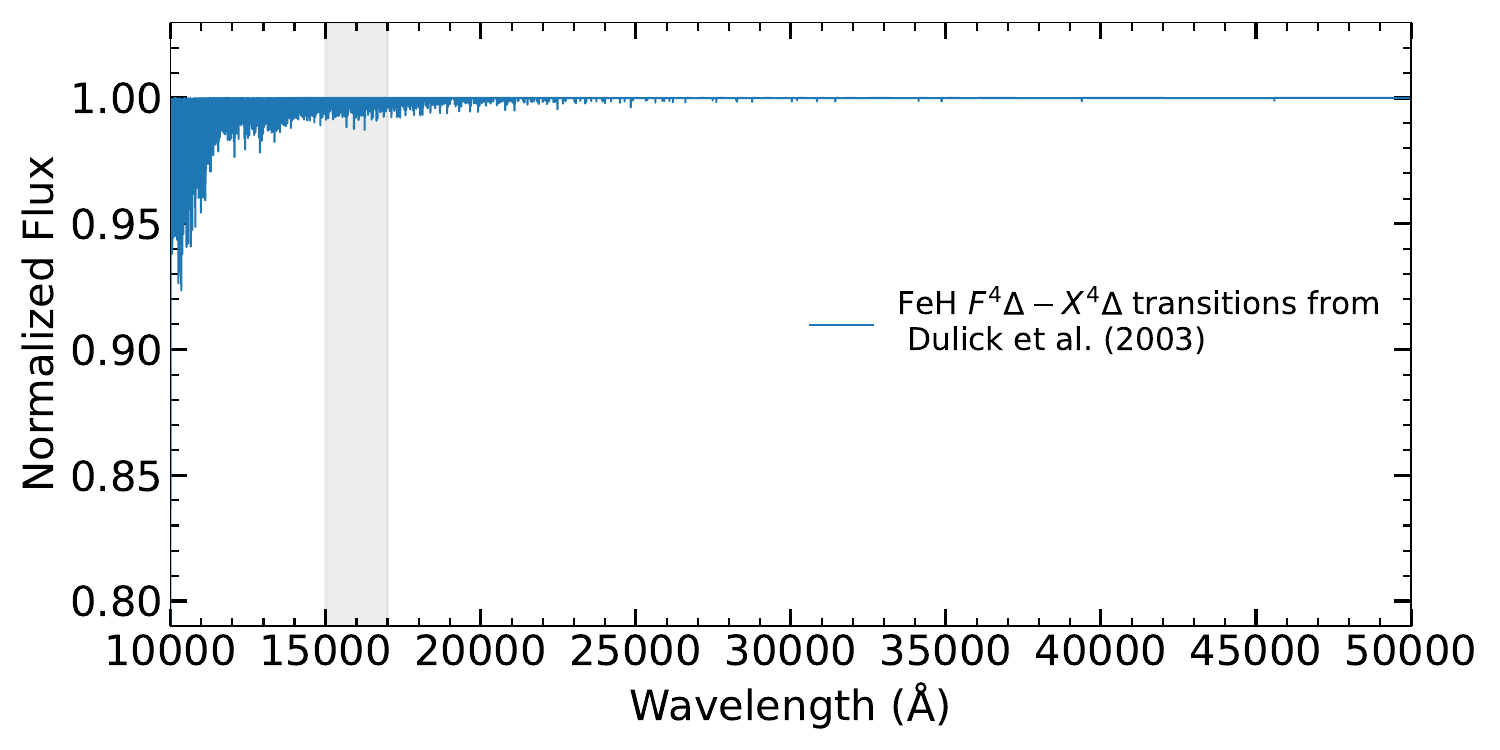}
    \caption{{\it Left panels:} top and bottom panels display log~$gf$ as a function of wavelength for \citet{Hargreaves2010} and \cite{Dulick2003} line lists, respectively. The orange circles correspond to the FeH diagnostic lines measured in this study, which are from the \citet{Hargreaves2010} line list. The shaded area demarcates the interval covered by APOGEE spectra. Note that the overwhelming majority of the \cite{Dulick2003} lines are very weak, with log~$gf<-2$.  
    {\it Right panels:} Synthetic spectra in the wavelength region between 1 -- 5 micron for the FeH \cite{Hargreaves2010} (top) and \cite{Dulick2003} (bottom) line lists, respectively, calculated for a \T~= 3900 K log $g$ = 5.0 dex and [Fe/H] = 0 dex MARCS model atmosphere.  Note that the \cite{Hargreaves2010} lines completely dominate the FeH opacity within the APOGEE spectral region. } 
    \label{fig:logfs}
\end{figure*}

Figure \ref{fig:logfs} displays log $gf$ values as a function of wavelength for the \cite{Hargreaves2010} (top left panel) and \cite{Dulick2003} (bottom left panel) line lists. The line list of \cite{Hargreaves2010} contains 6357 lines between $\lambda$1.5--2.0$\mu$m, whereas that by \cite{Dulick2003} includes 76,408 lines between $\lambda$1.0--5.0 $\mu$m. 
The shaded region indicates the wavelength range spanned by APOGEE spectra. The large, filled orange symbols in the top left panel mark the diagnostic FeH lines used in this study 
to derive M-dwarf fundamental parameters (Section \ref{sec:abundana}).
This figure illustrates the line densities and transition probabilities for FeH lines across the NIR for each of the FeH systems studied in \cite{Hargreaves2010} and \cite{Dulick2003}.
Important points to take away from Figure \ref{fig:logfs} include the distributions of the log~$gf$ values and line densities as functions of wavelength. The \cite{Dulick2003} list covering the F-X$^{4}\Delta$ transitions contains many more lines, with the highest density near $\lambda\sim$1.0-1.3 $\mu$m, although the majority of these FeH lines have log~$g$f values much smaller than -2 and are thus quite weak.  
In addition, as the wavelength increases, the distribution of the F-X$\Delta$ line excitation potentials increases significantly. Thus, the Boltzmann factor (e$^{-(\chi/kT)}$; where $\chi$, $k$, and $T$ are the excitation potential, the Boltzmann constant, and the absolute temperature, respectively) adds an additional term that greatly decreases their line strengths. Thus, FeH absorption due to the F-X$^{4}\Delta$ transitions (the Wing-Ford band) weakens rapidly beyond $\lambda\sim$1.0--1.3$\mu$m. The E-A$^{4}\Pi$ system studied by \cite{Hargreaves2010} begins to absorb at $\lambda\sim$1.58 $\mu$m and degrades redward from there.  
These FeH lines have maxima log $gf$$\sim$-1 values that are similar to the maxima of the F-X$^{4}\Delta$ lines, however, in the spectral region beyond $\lambda$1.6 $\mu$m the E-A$^{4}\Pi$ excitation potentials are all much lower, with $\chi\le$0.4 eV, so that within the APOGEE window ($\lambda\sim$1.5--1.7 $\mu$m) absorption from the E-A$^{4}\Pi$ lines are expected to dominate over the F-X$^{4} \Delta$ lines in M dwarfs.

As a result of the above combination of factors, one should expect that, within the APOGEE spectral window, the E-A$^{4}\Pi$ FeH lines produce the majority of iron hydride absorption in the spectra of M dwarfs. The right panels of Figure~\ref{fig:logfs} illustrate that expectation clearly, where it shows normalized synthetic spectra computed using Turbospectrum v19.1 (\citealp{AlvarezandPlez1997}, \citealp{Plez2012}) together with a MARCS model atmosphere (\citealp{Gustafsson2008}) for \T~= 3900 K, log~$g$ = 5.0 dex, and [Fe/H] = 0.0 dex. Spectra calculated adopting the \cite{Hargreaves2010} and \cite{Dulick2003} line lists are displayed on the top and bottom panels, respectively. The right bottom panel of Figure \ref{fig:logfs} shows that the \cite{Dulick2003} F-X$^{4}\Delta$ system produces stronger FeH lines near the Wing-Ford band, tapering off towards longer wavelengths. The \cite{Hargreaves2010} E-A$^{4}\Pi$ line list, on the other hand, only includes lines in the $\lambda\sim$1.5 -- 2 $\mu$m range. Within the APOGEE spectral region, it can be seen that the E-A$^{4}\Pi$ lines dominate the FeH absorption, as they are stronger than those by the F-X$^{4}\Delta$ by a factor of $\approx$~20.

\begin{figure*}[!t]
    \centering
    \includegraphics[height=0.17\textheight]{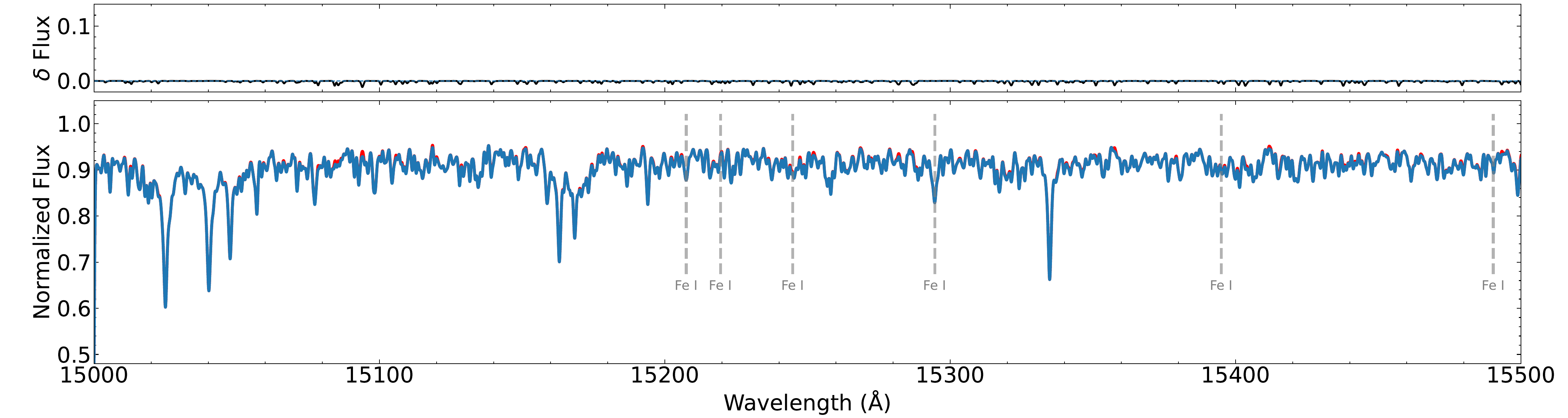}
    \includegraphics[height=0.17\textheight]{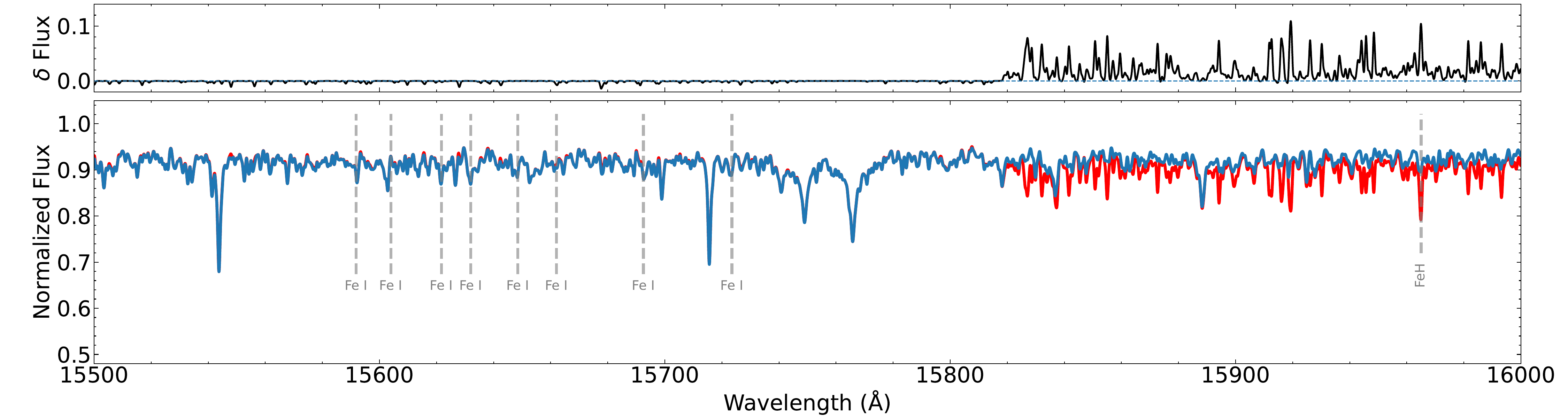}
    \includegraphics[height=0.17\textheight]{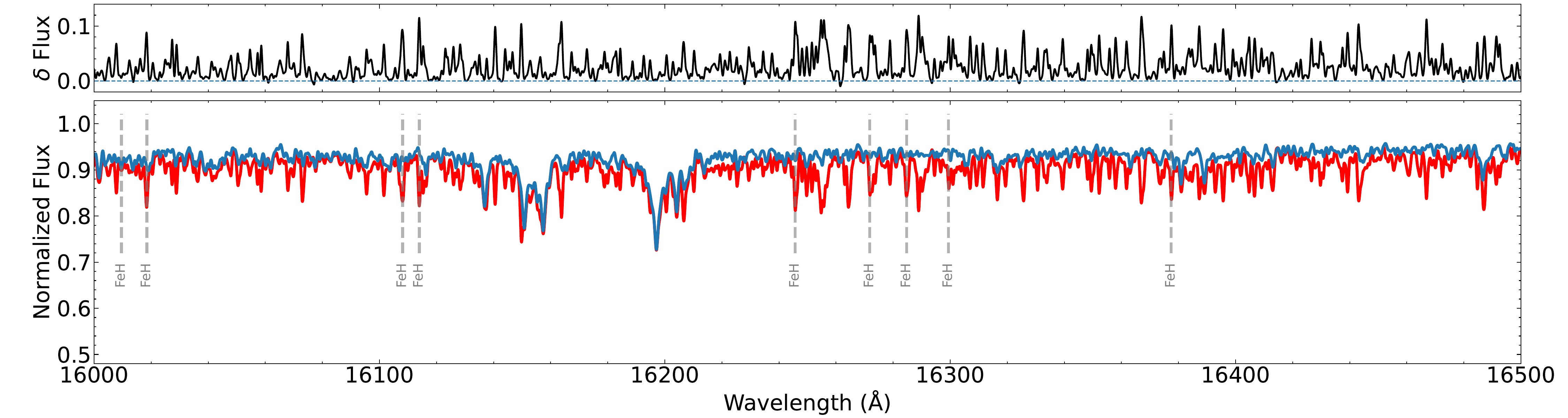}
    \includegraphics[height=0.17\textheight]{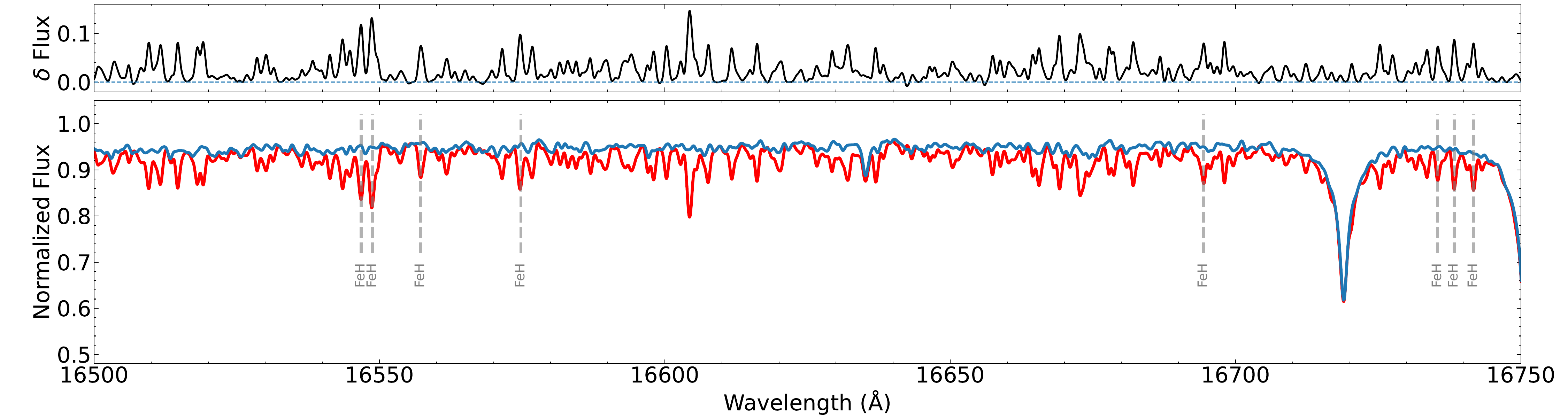}
    \includegraphics[height=0.17\textheight]{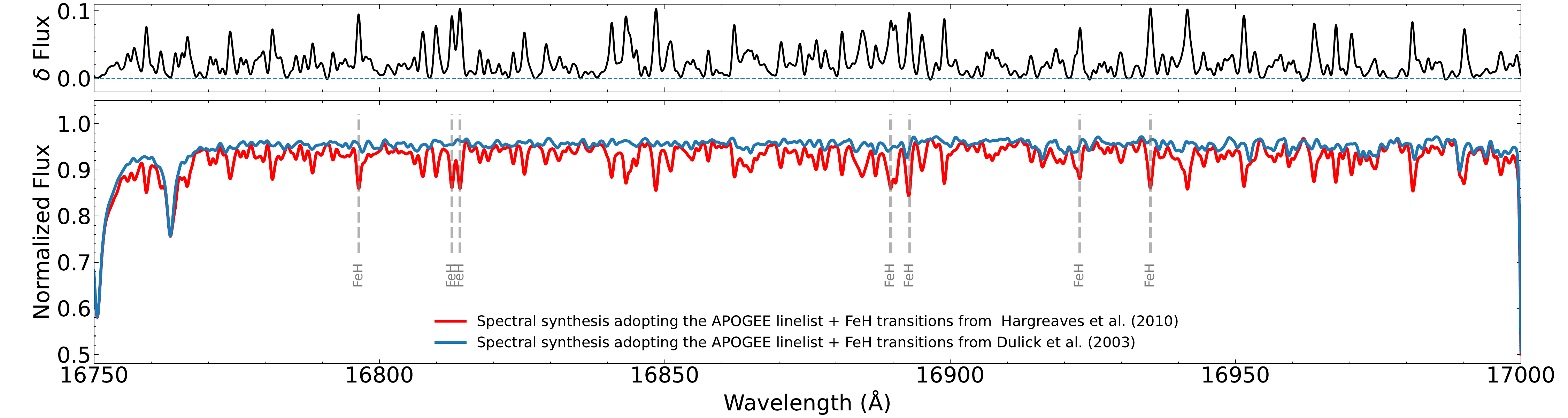}
    \caption{Synthetic spectra for \T~= 3500 K log $g$ = 5.0 dex and [Fe/H] = 0.0 dex, excluding the FeH line list from \cite{Hargreaves2010} (blue) or \cite{Dulick2003} (red). The impact of the \cite{Hargreaves2010} list for the E$^{4}\Pi$ – A$^{4}\Pi$ electronic transition is far more important in the APOGEE region than that of the \cite{Dulick2003} list for the F$^{4}\Delta$ – X$^{4}\Delta$.}
    \label{fig:compspec}
\end{figure*}

The impact that FeH has on M-dwarf spectra in the APOGEE region is illustrated in Figure \ref{fig:compspec}, where two synthetic spectra, convolved to the APOGEE resolution, are shown.  The spectra were synthesized using Turbospectrum with a MARCS model atmosphere ($T_{\rm eff}$ = 3500 K, log $g$ = 5.0 dex, and solar metallicity), and the APOGEE line list, but with the FeH lines from the F-X$^{4}\Delta$ system (\citealt{Dulick2003}) missing in one case (blue curve) and the FeH lines from the E-A$^{4}\Pi$ system (\citealt{Hargreaves2010}) missing in the other case (red curve). Figure \ref{fig:compspec} is divided into five main panels separated into wavelength intervals covering the APOGEE window. The bottom subpanel in each main panel shows the two synthetic spectra overplotted, while the top subpanels illustrate the subtracted differences between the two synthetic spectra. These spectra and their differences graphically demonstrate the importance of the E-A$^{4}\Pi$ FeH lines in fitting an M-dwarf spectrum redward of $\lambda$15820\AA\ and the relatively minor (if detectable at all) importance of the F-X$^{4}\Delta$ lines in this restricted APOGEE spectral window.  

\begin{deluxetable}{ccccccccc}
\tabletypesize{\scriptsize}
\setlength{\tabcolsep}{1.5pt}
\caption{Diagnostic FeH lines \label{tab:feh_lines}}
\tablehead{
\colhead{$\lambda$} & \colhead{$\chi$} & \colhead{log $gf$} & \colhead{D0} &
\colhead{Blend} & \colhead{$J_{\rm lower}$} & \colhead{Branch} &
\colhead{$\Omega$ \& Parity} & \colhead{$\lambda_{\rm star}$} \\
\colhead{(\AA)} & \colhead{(eV)} & \colhead{} & \colhead{} &
\colhead{} & \colhead{} & \colhead{} & \colhead{} & \colhead{(\AA)}
}
\startdata 
    15965.195 & 0.279 & -1.108 & 2.410 &  & & & & 15965.0\\
    16009.205 & 0.279 & -2.491 & 2.410 &  &  & & & 16009.6\\
    16018.420 & 0.279 & -1.078 & 2.410 &  &  & & &  16018.5\\
    16107.820 & 0.214 & -1.293 & 2.410 &  b1 &  9.5 & R & 1.5e & 16108.1 \\
    16108.337 & 0.231 & -1.250 & 2.410 &  b1 &  10.5 & R & 1.5e &\\
    16113.763 & 0.198 & -1.331 & 2.410 &  b2 &  8.5 & R & 1.5e & 16114.0 \\ 
    16114.049 & 0.279 & -1.282 & 2.410 &  b2 &  & & & \\
    16245.594 & 0.373 & -1.188 & 2.410 &  b3 &  16.5 & R & 1.5e & 16245.7 \\
    16245.746 & 0.142 & -1.409 & 2.410 &  b3 & 5.5 & R & 2.5e & \\
    16271.777 & 0.302 & -1.136 & 2.410 &  &  12.5 & R & 0.5e &  16271.8\\
    16284.665 & 0.229 & -1.274 & 2.410 &  &  11.5 & R & 2.5e &  16284.7\\
    16299.496 & 0.179 & -1.252 & 2.410 &   &  8.5 & R & 2.5f & 16299.4\\
    16377.403 & 0.344 & -1.203 & 2.410 &  b4 &  16.5 & R & 2.5e & 16377.4\\
    16377.688 & 0.279 & -1.439 & 2.410 &  b4 &  & & & \\
    16546.306 & 0.279 & -1.601 & 2.410 &  b5 &  & & & 16546.0\\
    16546.755 & 0.189 & -1.512 & 2.410 &  b5 &  6.5 & P & 0.5e & \\
    16546.898 & 0.133 & -1.632 & 2.410 &  b5 & 4.5 & P & 2.5e & \\
    16548.569 & 0.279 & -1.639 & 2.410 &  b6 &  & & & 16548.8 \\
    16548.891 & 0.305 & -1.336 & 2.410 &  b6 &  13.5 & R & 1.5f & \\
    16557.238 & 0.279 & -1.586 & 2.410 &  &  & & & 16557.2 \\
    16574.751 & 0.473 & -0.996 & 2.410 &  &  18.5 & R & 0.5e & 16574.8 \\
    16694.389 & 0.279 & -1.485 & 2.410 &  b7 &  & & & 16694.4 \\
    16694.502 & 0.279 & -1.687 & 2.410 &  b7 &  & & & \\
    16735.420 & 0.220 & -1.326 & 2.410 &  & 8.5 & P & 0.5e & 16735.4 \\
    16738.294 & 0.279 & -1.048 & 2.410 &  &  & & & 16738.3\\
    16741.657 & 0.165 & -1.341 & 2.410 &  &  7.5 & P & 2.5e & 16741.7\\
    16796.382 & 0.279 & -1.031 & 2.410 &  &  & & & 16796.4\\
    16812.687 & 0.279 & -1.041 & 2.410 &  &  & & & 16812.7\\
    16814.063 & 0.178 & -1.277 & 2.410 &  &  8.5 & P & 2.5e & 16814.1\\
    16889.575 & 0.194 & -1.239 & 2.410 &  &  9.5 & P & 2.5e & 16889.9\\
    16892.878 & 0.279 & -1.055 & 2.410 &  &  & & & 16892.9\\
    16922.746 & 0.251 & -1.259 & 2.410 &  &  11.5 & P & 1.5e & 16922.7\\
    16935.090 & 0.195 & -1.480 & 2.410 & b8 &  9.5 & P & 2.5f & 16935.1 \\
    16935.090 & 0.293 & -1.254 & 2.410 & b8 &  14.5 & Q & 2.5e & \\
\enddata
\end{deluxetable}

The FeH lines that were used as diagnostic lines in the determination of the stellar parameters in this study are presented in Table \ref{tab:feh_lines}. These FeH lines are all from the A$^{4}\Pi$--E$^{4}\Pi$ transition and were initially selected by \cite{Souto2017} and \cite{Souto2020}. Table \ref{tab:feh_lines} gives the air wavelengths of the FeH transitions from \cite{Hargreaves2010}, followed by the excitation potential (in eV) and the log $g$f of the lines. As discussed previously, \cite{Hargreaves2010} present a mixture of assigned and unassigned FeH transitions, and these are specified in Table \ref{tab:feh_lines} by their J value of the lower level, the branch, and the total angular momentum ($\Omega$) and Parity (e or f) of the transitions. Given the APOGEE spectral resolution, some of these FeH transitions are blended in the APOGEE spectra, and these lines are marked as `b1, b2, etc.' in Table \ref{tab:feh_lines}. The last column in the table has the measured wavelengths of the FeH features in the APOGEE spectra. 
The measured wavelengths from the star are stellar rest wavelengths and can be compared with those given by \cite{Hargreaves2010}. A wavelength comparison of the 16 apparently unblended features finds an average difference and standard deviation between $<\lambda_{\rm star}$ - $\lambda_{\rm Hargreaves}>$ = 0.00 $\pm$ 0.07 \AA.
Given that the nominal APOGEE resolution of $\sim$22,400 yields a resolution element of $\Delta\lambda\sim$0.7 \AA\ at $\lambda$16,000 \AA, a scatter in the measured wavelengths of $\pm$0.07 \AA\ suggests that the wavelengths obtained from \cite{Hargreaves2010} are reliable.

\section{Observational data}\label{sec:obsdata}

\subsection{The APOGEE spectra}\label{sec:spectral}

We used publicly available observational data from the Milky Way Mapper (MWM) Survey (data release 19; \citealp{sloandigitalskysurveyv2025}; \citealp{sdsscollaboration2025}) obtained using the APOGEE spectrograph.
APOGEE (\citealp{Wilson2010}) is a cryogenic multi-fiber survey instrument capable of observing up to 300 targets simultaneously in the NIR, spanning wavelengths from about 1.50 to 1.70 $\mu$m in the H band at a resolution $R = \lambda/\Delta\lambda \sim 22,500$. The survey uses two APOGEE spectrographs: one in the Northern Hemisphere, coupled to the 2.5 m Sloan Digital Sky Survey (SDSS) telescope \citep{Gunn2006} at Apache Point Observatory (APO), which also allows spectra to be collected via a secondary fiber feed from the 1m telescope at New Mexico State University \citep{Holtzman2015}; and another in the Southern Hemisphere, attached to the 2.5 m du Pont telescope \citep{Bowen:73} at Las Campanas Observatory (LCO).
Spectra were processed using the APOGEE pipeline, as detailed by \cite{Nidever_2015}, \cite{Holtzman_2018}, and \cite{Jönsson_2020}.

\subsection{The sample}\label{sec:spectra}

Our sample contains 36 main-sequence G- and M-type stars belonging to 18 wide binary systems. To select these systems, we cross-matched the APOGEE DR19 data with the wide-binary catalog of \citet{El_Badry}, compiled using Gaia EDR3 \citep{Gaia_eDR3} and including systems located at distances of about 1 kpc from the Sun. 
The raw cross-match yielded 1123 binary systems. For our sample, we selected only systems where the primary star is a G dwarf with \T~between 5000 K and 6000 K, and the secondary is an M dwarf with \T~between 3500 K and 3900 K. 
Furthermore, stellar spectra 
were required to have a signal-to-noise ratio (SNR) $\geq$ 100 to minimize the uncertainties in the derived parameters (\citealt{Jofre2019}).   
This selection criterion was adopted using non-calibrated, given in the FPARAM array of the APOGEE catalog, and SNR values from the APOGEE DR19 ASPCAP pipeline (\citealt{Perez2016}).
We note that binaries having K-type dwarf primaries have not been selected because the APOGEE abundances of K-dwarfs have been shown to have systematic uncertainties, being systematically lower than expected due to the selection of diagnostic lines in the ASPCAP pipeline that was optimized for the analysis of red-giant stars (see discussion in \citealt{Grilo2024}). 

\begin{figure}[ht!]
\centering
    \includegraphics[width=0.44\textwidth]{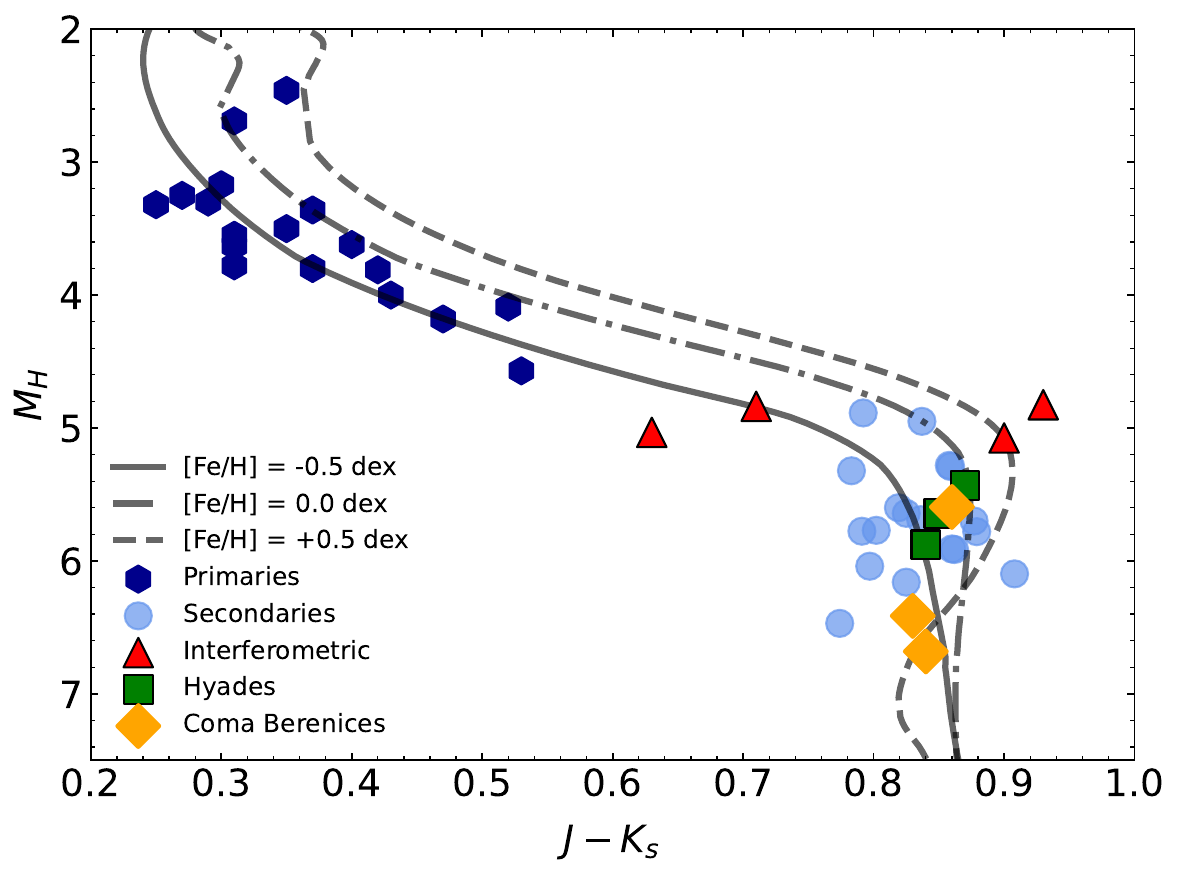}
    \includegraphics[width=0.45\textwidth]{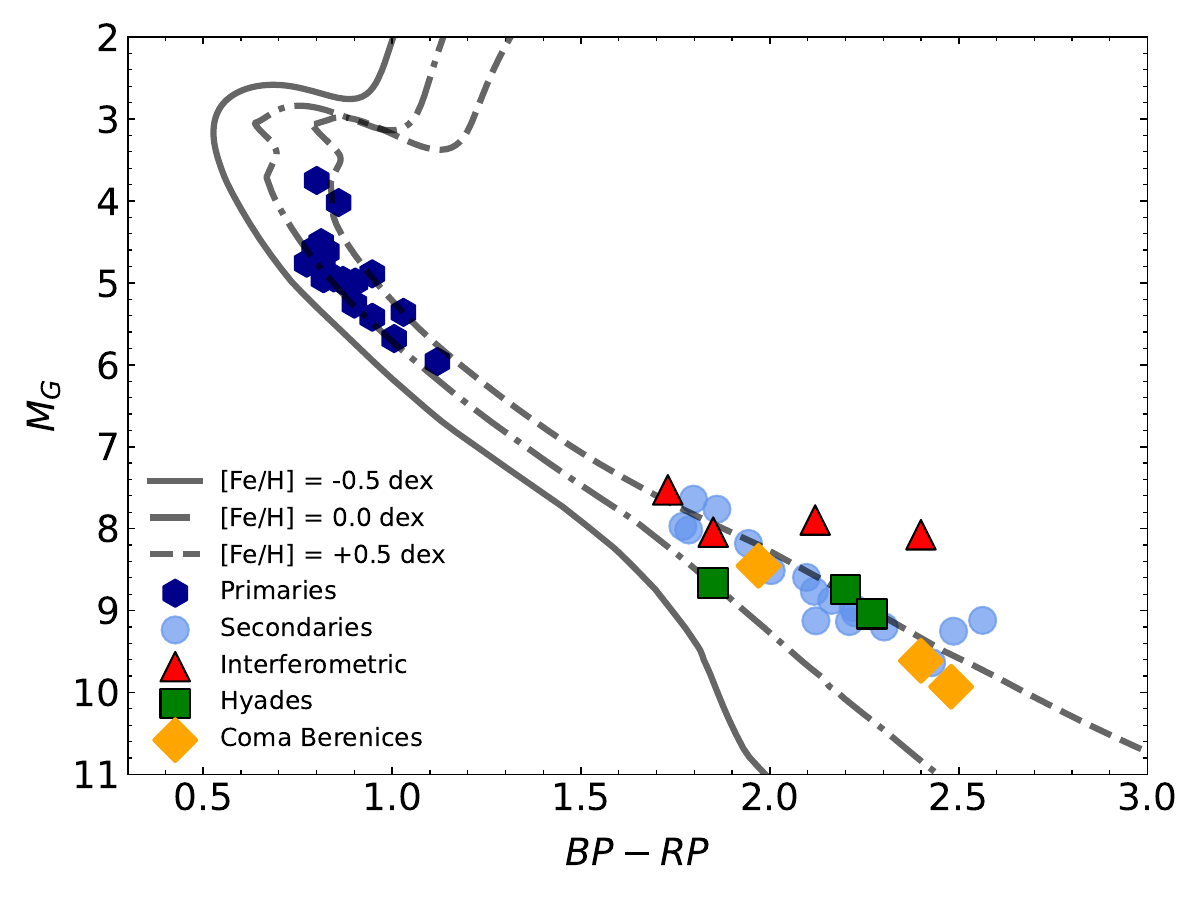}
    \caption{Color–magnitude diagram of the sample studied in this work. The top/bottom panel shows the 2MASS/Gaia photometry. Primary stars are plotted as dark blue hexagons, and secondary stars as light blue circles, while the interferometric, Hyades, and Coma Berenices stars are shown as red triangles, green squares, and orange diamonds, respectively. Three MIST isochrones of solar age (4.5 Gyr) with different initial metallicities (-0.5, 0.0, and +0.5 dex, respectively) are plotted.}
\label{fig:CMD}
\end{figure}

After application of the selection criteria for \T~and SNR, our sample consisted of 25 binary systems. This number was then reduced to 18 after removing those systems showing double lines in their spectra and classified as eclipsing binaries, stars with high $v \sin i$ ($>$ 10 km.s$^{-1}$), and those with poor telluric correction. Our final binary sample is presented in Table \ref{stellaratm} and in the top panel of Figure \ref{fig:CMD}, where the stars are distributed on a color-magnitude diagram (CMD) of M$_{\rm H}$ versus ($J - K_s$) using 2MASS photometry (\citealp{Skrutskie_2006}), while in the bottom panel of the figure we show the CMD using Gaia photometry (\citealp{Gaia_eDR3}).
Darker blue hexagons represent the G-type primary stars, and light blue circles indicate the M-type secondary stars. We also include in this figure three MIST isochrones (\citealt{Choi2016}) with solar age (4.5 Gyr) and different metallicities: -0.5 dex (black solid), 0.0 dex (black dot-dashed), and +0.5 dex (black dashed) for reference.
No reddening correction was applied to the photometry because all stars lie in the solar vicinity (d $<$ 210 pc).



\movetabledown=2in
\begin{deluxetable*}{cccccccc}
\tabletypesize{\scriptsize}
\tablecaption{The Sample and Stellar Parameters \label{stellaratm}}
\tablehead{
\colhead{2MASS~ID} & \colhead{$J$} & \colhead{$K_s$} & \colhead{$H$} & \colhead{SNR} & \colhead{log~$g$} & \colhead{\T} & \colhead{[Fe/H]}\\
\colhead{} & \colhead{} & \colhead{} & \colhead{} & \colhead{} & \colhead{(dex)} & \colhead{(K)} & \colhead{(dex)} }
\startdata
     \textbf{Primaries} \\
     2M12413348+4105187 & 9.64 & 9.17 & 9.21 & 392 & 4.36 & 5223 & -0.29 \\
     2M06315141+0038594 & 8.16 & 7.77 & 7.80 & 200 & 4.31 & 5547 & -0.24 \\
     2M08073608+3040585 & 7.55 & 7.20 & 7.25 & 774 & 4.34 & 5780 & -0.03 \\
     2M08092593+5203506 & 8.60 & 8.12 & 8.23 & 487 & 4.32 & 5374 & 0.21 \\
     2M08490883+1122497 & 9.83 & 9.49 & 9.58 & 432 & 4.40 & 5816 & -0.11 \\
     2M14045868+0156589 & 7.51 & 7.06 & 7.19 & 337 & 4.41 & 5539 & 0.00 \\
     2M19000197+5657420 & 7.44 & 7.09 & 7.19 & 240 & 4.21 & 5874 & 0.06 \\
     2M04551276+0347124 & 8.66 & 8.25 & 8.35 & 271 & 4.43 & 5712 & -0.17 \\
     2M08285922-0222368 & 8.61 & 8.30 & 8.61 & 243 & 4.51 & 5913 & -0.11 \\
     2M07340884-0553426 & 9.11 & 8.63 & 9.11 & 182 & 4.44 & 5462 & -0.07 \\
     2M04565146+0435101 & 8.75 & 8.22 & 8.75 & 325 & 4.47 & 4952 & 0.08 \\
     2M19110056-0741050 & 8.21 & 7.84 & 8.21 & 400 & 4.47 & 5612 & 0.10 \\
     2M00251041+6826409 & 10.40	& 10.01 & 10.09 & 101 & 4.44 & 5594 & 0.15 \\
     2M11565777+2022504 & 11.03	& 10.52 & 10.59 & 165 & 4.48 & 5181 & -0.03 \\
     2M22132300+1356457 & 9.25 & 8.88 & 8.95 & 394 & 4.43 & 5822 & 0.04 \\
     2M09023053$-$0040415 & 8.30 & 7.78 & 8.30 & 1538 & 4.37 & 5171 & 0.29 \\
     2M02045009$-$0015433 & 9.21 & 8.79 & 8.90 & 445 & 4.23 & 5740 & 0.31 \\
     2M03040962+6142212 & 5.39 & 5.02 & 5.11 & 203 & 4.35 & 5657 & -0.25 \\
     \hline
     \textbf{Secondaries}\\
     2M12414006+4103080 & 12.08	& 11.29 & 11.48 & 108 & 4.50  & 3751 & -0.20 \\
     2M06312373+0036445 & 11.07	& 10.25 & 10.46 & 210 & 4.72  & 3746 & -0.19 \\
     2M08083496+3047575 & 11.84 & 11.00 & 11.16 & 221 & 4.69  & 3803 & -0.07 \\
     2M08092559+5202190 & 11.15	& 10.28 & 10.51 & 143 & 4.54  & 3585 & 0.12 \\
     2M08485678+1124111 & 10.12	& 9.26 & 9.48 & 100 & 4.50 & 3824 & 0.04 \\
     2M14045583+0157230 & 10.12	& 9.26 & 9.48 & 250 & 4.72 & 3684 & 0.01 \\
     2M19000114+5657430 & 11.15	& 10.36 & 10.50 & 206 & 4.50 & 3786 & -0.01 \\
     2M04551359+0347227 & 10.64 & 9.86 & 10.03 & 169 & 4.50 & 4005 & -0.15 \\
     2M08290608$-$0222560 & 10.80 & 9.94 & 10.10 & 100 & 4.50 & 4100 & -0.03 \\
     2M07340693$-$0553127 & 10.48 & 9.68 & 9.81 & 104 & 4.50 & 3983 & -0.10 \\
     2M04564625+0435309 & 10.90 & 9.99 & 10.27 & 116 & 4.81 & 3651 & 0.11 \\
     2M19111634$-$0738535 & 10.59 & 9.77 & 10.00 & 171 & 4.50 & 3915 & 0.05 \\
     2M00251625+6825399 & 12.50 & 11.64 & 11.82 & 108 & 4.58 & 4017 & 0.13 \\
     2M11565402+2022392 & 12.82 & 12.01 & 12.18 & 108 & 4.50 & 3952 & -0.15 \\
     2M22132300+1356336 & 12.09 & 11.22 & 11.39 & 116 & 4.60 & 3737 & 0.11 \\
     2M09025200$-$0040368 & 10.76 & 9.90 & 10.12 & 417 & 4.77 & 3613 & 0.17 \\
     2M02044840$-$0013563 & 11.84 & 11.00 & 11.16 & 120 & 4.50 & 4039 & 0.34 \\
     2M03044335+6144097 & 8.87 & 8.10 & 8.32 & 160 & 4.69 & 3591 & -0.04 \\
     \hline
     \textbf{$^a$Interferometric/$^{b,c}$Open cluster}\\
     2M05312734$-$0340356$^a$ & 4.83 & 3.90 & 4.05 & 197 & 4.84 & 3928 & 0.16 \\
     2M09142485+5241118$^a$ & 4.77 & 4.14 & 4.04 & 153 & 4.78 & 4030 & -0.03 \\
     2M09142298+5241125$^a$ & 4.89 & 3.99 & 3.98 & 144 & 5.00 & 4041 & -0.05 \\
     2M10112218+4927153$^a$ & 3.97 & 3.26 & 3.27 & 144 & 4.50 & 4234 & 0.05 \\
     2M04295572+1654506$^b$ & 9.52 & 8.65 & 8.65 & 514 & 4.72 & 3796 & 0.09 \\
     2M04291097+2614484$^b$ & 9.68 & 8.83 & 9.04 & 297 & 4.50 & 3805 & 0.06 \\
     2M03591417+2202380$^b$ & 9.54 & 8.70 & 8.91 & 226 & 4.88 & 3767 & 0.09 \\
     2M12255421+2651387$^c$ & 11.98 & 11.14 & 11.39 & 290 & 5.00 & 3501 & 0.06 \\
     2M12250262+2642382$^c$ & 11.62 & 10.79 & 11.03 & 385 & 4.98 & 3564 & 0.08 \\
     2M12241121+2653166$^c$ & 10.92 & 10.06 & 10.27 & 507 & 4.82 & 3912 & -0.07 \\
\enddata
\tablecomments{Effective temperatures and log g values for the G-dwarf primaries are from MWM DR19. (a) Star with measured angular diameter; (b) Hyades open cluster member; (c) Coma Berenices open cluster member. The full table is available in machine-readable form.}
 \end{deluxetable*}

We also analyzed 10 other benchmark M dwarfs previously studied in the literature (shown in Figure \ref{fig:CMD}), including four interferometric early-type M dwarfs (triangles) from \cite{Boyajian2012}, three M dwarf members of the Coma Berenices open cluster (diamonds) from \cite{Souto2021}, and three from the Hyades open cluster (squares) studied in \cite{Wanderley2023} and \cite{Vilar2025}. 

\section{Analysis} 
\label{sec:abundana}

We used the BACCHUS wrapper \citep{Masseron2016} to generate spectral syntheses for this work. BACCHUS computes synthetic spectra using the Turbospectrum radiative transfer code (\citealp{AlvarezandPlez1997}; \citealp{Plez2012}), 1D local thermodynamic equilibrium (LTE) plane-parallel MARCS model atmospheres \citep{Gustafsson2008}, and the APOGEE line list \citep{Smith2021}. It performs synthesis on the fly and compares the results directly with the observed spectrum. 
BACCHUS can operate in two modes: fully automatic and semi-automatic. For this analysis, we employed the semi-automatic mode to allow minor adjustments to the pseudo-continuum, abundances, radial velocities, or line broadening as needed.
We adopted a microturbulence velocity ($\xi$) of 1.00 $\pm$ 0.25 km.s$^{-1}$ \citep{Souto2017} in the calculations of synthetic spectra.

\begin{figure*}[!t]
    \centering
    \includegraphics[width=0.44\textwidth]{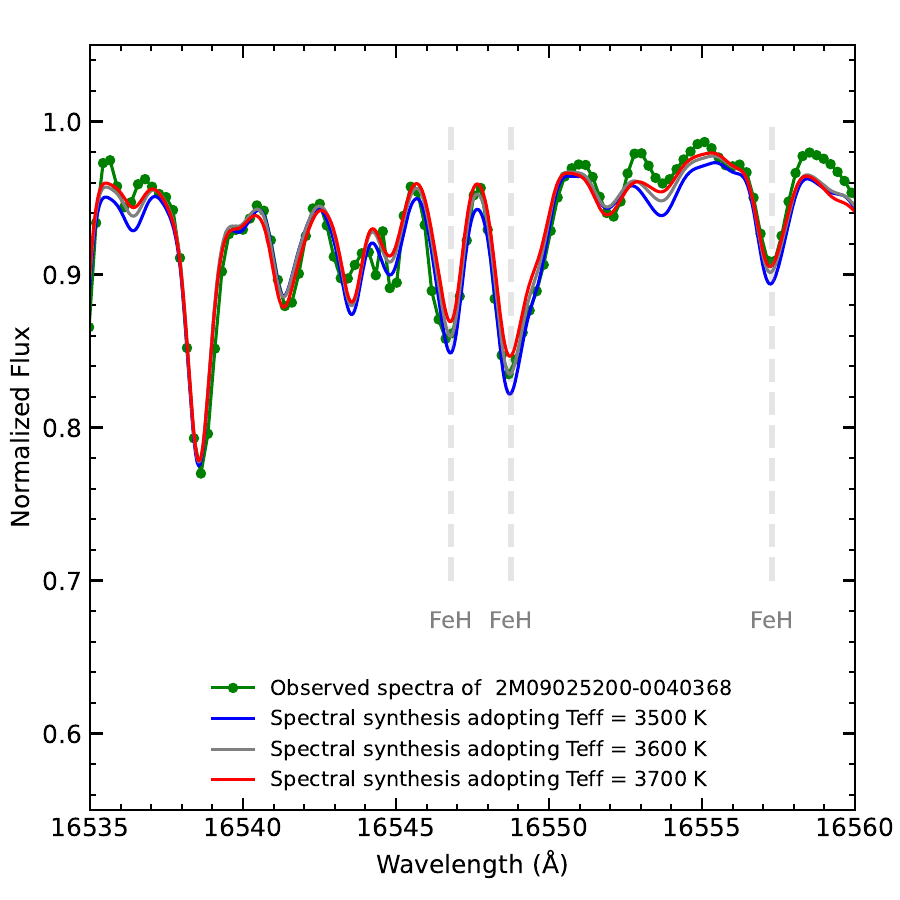}
    \includegraphics[width=0.44\textwidth]{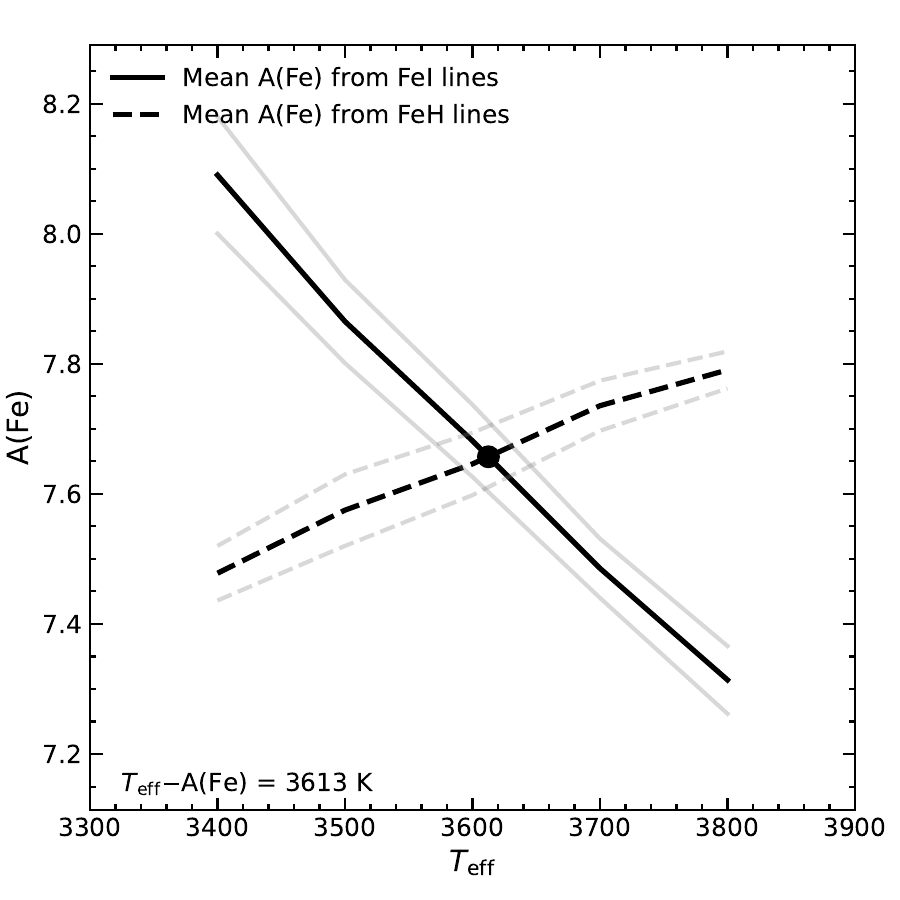}
    \includegraphics[width=0.44\textwidth]{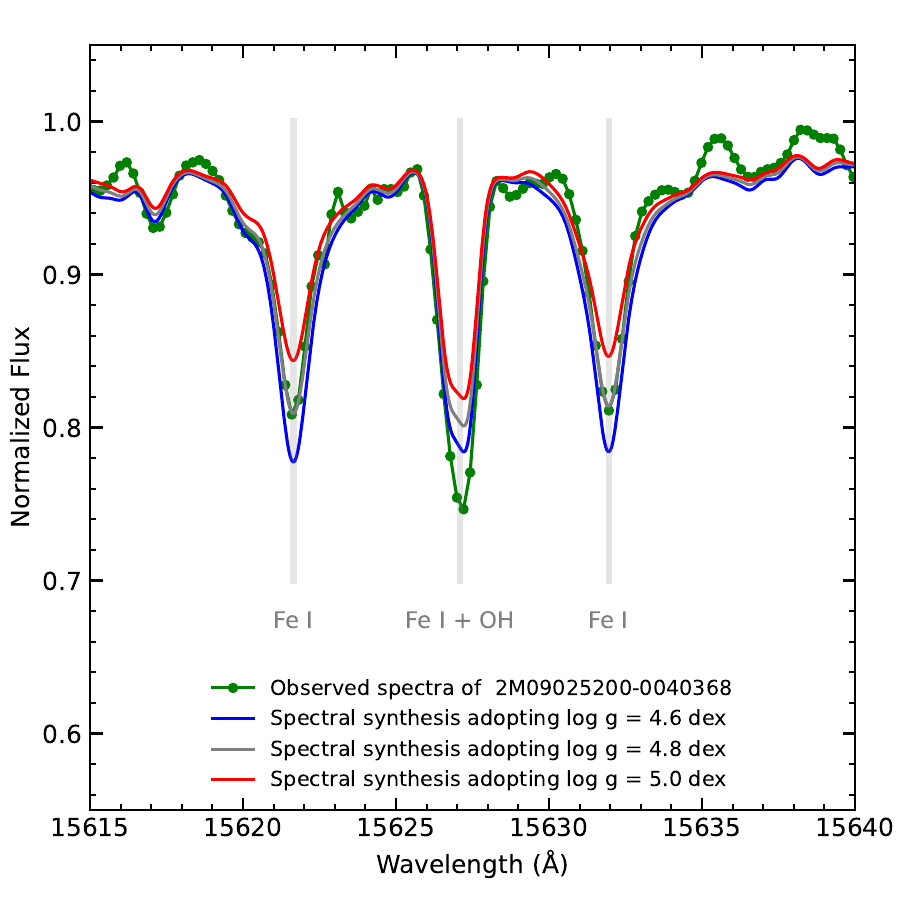}
    \includegraphics[width=0.44\textwidth]{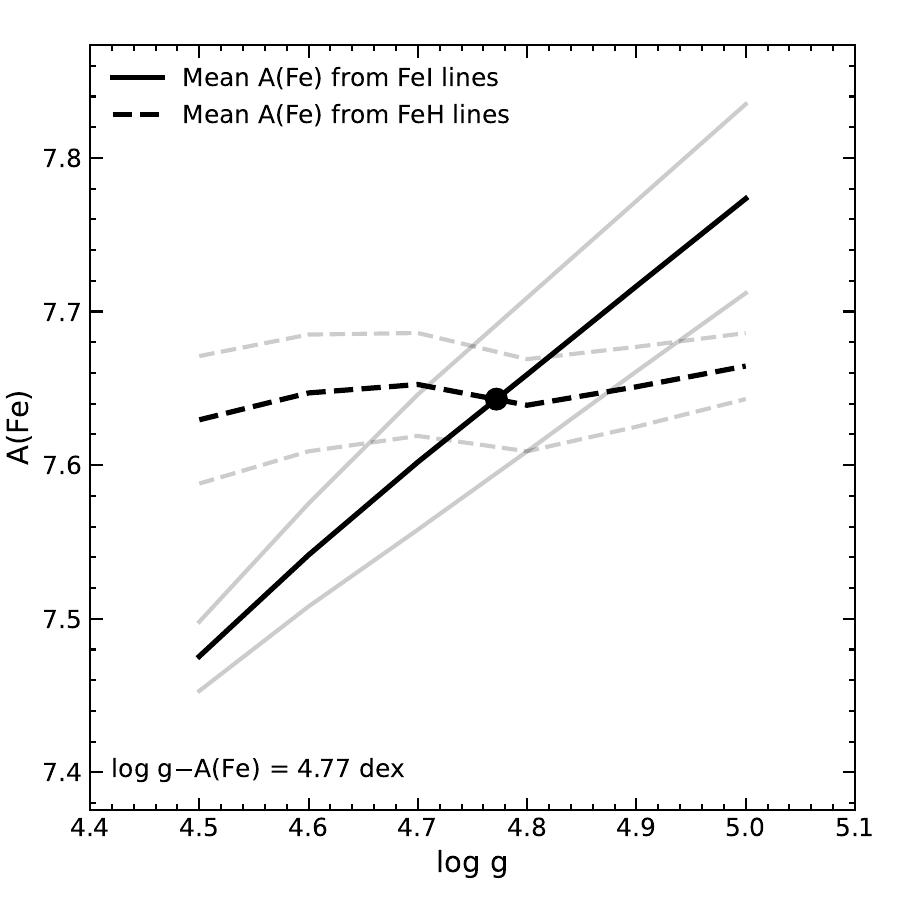}
    \caption{{\it Left panels:}  portions of APOGEE spectra for the M dwarf 2M09025200-0040368 (green dotted line). The blue, gray, and red solid lines represent synthetic spectra computed for \T~= 3500, 3600, 3700 K (top panel) and log $g$ = 4.6, 4.8, 5.0 dex (bottom panel). 
    {\it Right panels:} illustration of the \T~(top panel) and log~$g$ (bottom panel) determinations. The solid and dashed black lines correspond to iron abundances derived from Fe I and FeH lines. The black-filled circles represent our \T-A(Fe) and log $g$-A(Fe) pairs, indicating an intersection point where the abundance indicators agree and represent our derived \T~and log $g$, respectively.}
    \label{fig:specteff}
\end{figure*}

\subsection{Effective Temperatures and Surface Gravities}

We derived stellar parameters for the M dwarfs by enforcing consistency between the iron abundances obtained from Fe I and FeH lines. We follow the procedure of \cite{Souto2018}.
The method consists of iteratively varying first \T\ and then log g, 
while deriving the corresponding iron abundances. Here, we consider the \T--A(Fe) and log $g$--A(Fe) pairs only in order to more directly assess the reliability of the derived atmospheric parameters and Fe abundances from FeH lines. In this approach, we determine a \T~that yields consistent abundances from both indicators (Fe I and FeH).

Figure \ref{fig:specteff} exemplifies the methodology, where the top-left panel displays a portion of the observed spectrum (dotted green line) of the M dwarf 2M09025200-0040368, along with synthetic spectra computed for three different effective temperatures: \T~= 3500 K (blue), 3600 K (gray), and 3700 K (red). The bottom-left panel follows the same format, but presents synthetic spectra computed for different surface gravity values, with log $g$ = 4.6 (blue), 4.8 (black), and 5.0 (red) dex. In both panels, the Fe I and FeH lines are marked.  
The right panels illustrate the determination of \T~(top-right panel) and log $g$ (bottom-right panel). Black solid lines correspond to abundances derived from Fe I transitions, while black dashed lines represent those from FeH transitions; the filled circle marks the consistent \T-A(Fe) and log $g$-A(Fe) pairs.  
The gray lines represent edges limited by the iron abundance uncertainties (\citealt{Melo2024}), and these are used to derive the parameter uncertainties for each star.
As shown in Figure \ref{fig:specteff}, the Fe I and FeH lines exhibit different sensitivities to variations in \T: Fe I lines are more responsive, showing significant changes, whereas FeH lines exhibit smaller variations. The trend of A(Fe) with surface gravity for FeH is similar to that with \T, while the behaviors of A(Fe) versus \T~ and log $g$ are opposite for Fe I. 

The same methodology adopted here to derive effective temperatures and surface gravities for M dwarfs discussed above does not apply to the G dwarfs, as in hotter dwarf stars (\T~$>$ 3950 K), H$_2$O lines become too weak to be effectively utilized. Similarly, FeH (or Fe II) lines are not available for FG-type dwarfs (4300 K $<$ \T~$<$ 6000 K) in the APOGEE region, making it impossible to apply this methodology to hotter stars. In this study, we adopted the following stellar parameters for the primary stars: raw \T~values from ASPCAP DR19 and calibrated ASPCAP DR19 log $g$ values. 
The ASPCAP derived $T_{effs}$ are expected to be precise, with uncertainties usually lower than 50 K for solar-like stars and red giants (\citealt{Holtzman_2018}).

\subsection{Metallicities}

We determined iron abundances by measuring FeH lines and Fe I lines for the M dwarfs, and Fe I lines for our sample of G-type dwarfs, with the selection of 77 Fe I lines for the G-types taken from \cite{Grilo2024} and for the M-types from \cite{Melo2024}. 
The Fe abundances were derived from best fits to the lowest chi-squared values from BACCHUS between the observed and synthetic spectral lines, and by checking that satisfactory visual fits were obtained.  
The Fe abundances for each measured line in this study are presented in Table \ref{tab:feh_line_abundances} for the G dwarfs, and in Table \ref{tab:mdwarf_line_abundances_landscape} for the M dwarfs in our sample, 
with the mean iron abundance and standard deviation for each star given at the bottom of these tables. 
Solar reference values are taken from \citet{Asplund2021}.


\begin{table*}
\centering
\scriptsize
\setlength{\tabcolsep}{3.5pt}
\caption{Individual line Fe I abundances for the G-Dwarf stars}
\label{tab:feh_line_abundances}

\begin{minipage}{\textwidth}
\centering
\vspace{4pt}

\begin{tabular}{lccccccccccccccccccc}
\hline\hline
$\lambda$ (\AA) &
5187 & 8594 & 0585 & 3506 & 2497 & 6589 & 7420 & 7124 &
2368 & 3426 & 5101 & 1050 & 6409 & 2504 & 6457 & 0415 & 5433 & 2212 \\
\hline
15194.5  & 7.24 & 7.30 & \ldots & 7.76 & 7.55 & 7.51 & 7.68 & 7.41 & 7.38 & 7.44 & 7.56 & 7.68 & 7.39 & 7.28 & 7.28 & 7.75 & \ldots & 7.43 \\
15207.5  & 7.16 & 7.19 & 7.43   & 7.69 & 7.42 & 7.51 & 7.48 & 7.34 & 7.47 & 7.49 & 7.66 & 7.62 & 7.71 & 7.54 & 7.53 & 7.83 & 7.72   & 7.14 \\
15219.6  & \ldots & \ldots & \ldots & \ldots & \ldots & \ldots & \ldots & \ldots & \ldots & \ldots & \ldots & \ldots & \ldots & \ldots & \ldots & \ldots & \ldots & \ldots \\
15224.5  & 7.13 & 7.16 & 7.42   & 7.69 & 7.42 & 7.46 & 7.52 & 7.34 & 7.25 & 7.44 & 7.56 & 7.56 & 7.61 & 7.43 & 7.48 & 7.78 & 7.81   & 7.17 \\
15239.9 & 7.09 & 7.06 & 7.43 & 7.70 & 7.20 & 7.31 & 7.36 & 7.24 & 7.39 & 7.34 & 7.50 & 7.47 & 7.51 & 7.36 & 7.37 & 7.72 & 7.73 & 7.34 \\
$\vdots$ & $\vdots$ & $\vdots$ & $\vdots$ & $\vdots$ & $\vdots$ & $\vdots$ & $\vdots$ & $\vdots$ & $\vdots$ & $\vdots$ & $\vdots$ & $\vdots$ & $\vdots$ & $\vdots$ & $\vdots$ & $\vdots$ & $\vdots$ & $\vdots$ \\
16665.4  & 7.20 & 7.10 & 7.49   & 7.70 & 7.41 & 7.51 & 7.52 & 7.29 & 7.38 & 7.42 & 7.55 & 7.58 & 7.68 & 7.48 & 7.56 & 7.73 & 7.76   & 7.21 \\
\hline
\multicolumn{19}{l}{\textit{Summary statistics}} \\
\hline
$A(\mathrm{Fe})$ mean    & 7.17 & 7.22 & 7.43 & 7.67 & 7.38 & 7.47 & 7.52 & 7.30 & 7.35 & 7.40 & 7.53 & 7.55 & 7.61 & 7.44 & 7.49 & 7.73 & 7.76 & 7.23 \\
$A(\mathrm{Fe})$ median  & 7.17 & 7.22 & 7.42 & 7.66 & 7.37 & 7.46 & 7.52 & 7.29 & 7.36 & 7.40 & 7.52 & 7.55 & 7.62 & 7.43 & 7.48 & 7.75 & 7.76 & 7.20 \\
$A(\mathrm{Fe})$ std     & 0.07 & 0.08 & 0.05 & 0.05 & 0.07 & 0.07 & 0.07 & 0.06 & 0.08 & 0.08 & 0.07 & 0.07 & 0.07 & 0.09 & 0.07 & 0.10 & 0.04 & 0.11 \\
$[\mathrm{Fe/H}]$ mean   & $-$0.29 & $-$0.24 & $-$0.03 & 0.21 & $-$0.08 & 0.01 & 0.06 & $-$0.16 & $-$0.11 & $-$0.06 & 0.07 & 0.09 & 0.15 & $-$0.02 & 0.03 & 0.27 & 0.30 & $-$0.23 \\
$[\mathrm{Fe/H}]$ median & $-$0.29 & $-$0.24 & $-$0.04 & 0.20 & $-$0.09 & 0.00 & 0.06 & $-$0.17 & $-$0.10 & $-$0.06 & 0.06 & 0.09 & 0.16 & $-$0.03 & 0.02 & 0.29 & 0.30 & $-$0.26 \\
\hline
\end{tabular}

\vspace{2pt}
{\footnotesize\textbf{Note}. Column headers give the last four digits of each star’s 2MASS designation. Missing measurements are indicated by . . . . Only 6 of 77 Fe I lines shown. Solar reference \cite{Asplund2021}. This table is available in its entirety in machine-readable form.}
\end{minipage}

\end{table*}

\begin{table*}
\centering
\scriptsize
\setlength{\tabcolsep}{3.5pt}
\caption{Individual line Fe I and FeH abundances for the M-Dwarf stars}
\label{tab:mdwarf_line_abundances_landscape}

\begin{minipage}{\textwidth}
\centering
\vspace{2pt}

\begin{tabular}{llccccccccccccccccc}
\hline\hline
$\lambda$ (\AA) & Species &
3080 & 6445 & 7575 & 2190 & 4111 & 7230 & 7430 & 7227 &
2560 & 3127 & 5309 & 8535 & 5399 & 2392 &
$\cdots$ & 3166 \\
\hline
\multicolumn{18}{l}{\textit{Fe\,\textsc{i} lines}} \\
\hline
15207.5  & Fe\,\textsc{i} & 7.29 & 7.41 & 7.44 & 7.49 & 7.56 & 7.49 & 7.44 & 7.39 & 7.42 & 7.36 & 7.38 & 7.48 & 7.67 & 7.39 & $\cdots$ & 7.40 \\
15219.5  & Fe\,\textsc{i} & \ldots & \ldots & \ldots & \ldots & \ldots & \ldots & \ldots & \ldots & \ldots & \ldots & \ldots & \ldots & \ldots & \ldots & $\cdots$ & \ldots \\
15244.8  & Fe\,\textsc{i} & \ldots & \ldots & 7.52 & \ldots & 7.36 & \ldots & 7.44 & 7.38 & 7.48 & 7.47 & \ldots & 7.47 & 7.67 & 7.30 & $\cdots$ & \ldots \\
15294.6  & Fe\,\textsc{i} & 7.24 & 7.07 & 7.31 & 7.52 & 7.50 & 7.43 & 7.39 & 7.24 & 7.26 & 7.26 & 7.56 & 7.37 & 7.56 & 7.25 & $\cdots$ & 7.35 \\
$\vdots$ & $\vdots$ & $\vdots$ & $\vdots$ & $\vdots$ & $\vdots$ & $\vdots$ & $\vdots$ & $\vdots$ & $\vdots$ & $\vdots$ & $\vdots$ & $\vdots$ & $\vdots$ & $\vdots$ & $\vdots$ & $\vdots$ & $\vdots$ \\
15723.5  & Fe\,\textsc{i} & 7.31 & 7.19 & 7.32 & 7.71 & 7.69 & \ldots & 7.48 & 7.37 & 7.48 & 7.39 & 7.70 & 7.58 & 7.64 & 7.34 & $\cdots$ & 7.30 \\
\hline
\multicolumn{18}{l}{\textit{FeH lines}} \\
\hline
15965.0  & FeH & \ldots & \ldots & \ldots & 7.53 & \ldots & \ldots & \ldots & \ldots & \ldots & \ldots & \ldots & \ldots & \ldots & \ldots & $\cdots$ & \ldots \\
16009.6  & FeH & \ldots & \ldots & \ldots & \ldots & \ldots & \ldots & \ldots & \ldots & \ldots & \ldots & \ldots & \ldots & \ldots & \ldots & $\cdots$ & \ldots \\
16018.5  & FeH & \ldots & \ldots & \ldots & 7.38 & \ldots & \ldots & 7.46 & \ldots & \ldots & \ldots & 7.62 & 7.48 & \ldots & 7.39 & $\cdots$ & 7.44 \\
16108.1  & FeH & 7.25 & \ldots & 7.41 & 7.70 & \ldots & 7.49 & \ldots & \ldots & \ldots & \ldots & 7.61 & 7.61 & 7.60 & 7.31 & $\cdots$ & \ldots \\
$\vdots$ & $\vdots$ & $\vdots$ & $\vdots$ & $\vdots$ & $\vdots$ & $\vdots$ & $\vdots$ & $\vdots$ & $\vdots$ & $\vdots$ & $\vdots$ & $\vdots$ & $\vdots$ & $\vdots$ & $\vdots$ & $\vdots$ & $\vdots$ \\
16935.1  & FeH & \ldots & \ldots & 7.36 & \ldots & \ldots & \ldots & \ldots & 7.30 & \ldots & \ldots & 7.54 & \ldots & \ldots & \ldots & $\cdots$ & \ldots \\
\hline
\multicolumn{18}{l}{\textit{Summary statistics}} \\
\hline
$A(\mathrm{FeH})$ mean   & & 7.25 & 7.32 & 7.39 & 7.57 & 7.47 & 7.51 & 7.47 & 7.30 & 7.45 & 7.37 & 7.59 & 7.53 & 7.58 & 7.31 & $\cdots$ & 7.42 \\
$A(\mathrm{FeH})$ std    & & 0.09 & 0.11 & 0.06 & 0.09 & 0.11 & 0.07 & 0.10 & 0.11 & 0.10 & 0.09 & 0.06 & 0.05 & 0.10 & 0.08 & $\cdots$ & 0.05 \\
$[\mathrm{FeH/H}]$ mean  & & $-$0.21 & $-$0.14 & $-$0.07 & 0.11 & 0.01 & 0.05 & 0.01 & $-$0.16 & $-$0.01 & $-$0.09 & 0.13 & 0.07 & 0.12 & $-$0.15 & $\cdots$ & $-$0.04 \\
\hline
\end{tabular}

\vspace{2pt}
{\footnotesize\textbf{Note}. Column headers give the last four digits of each star’s 2MASS designation. Only 15 of 28 stars shown. Missing measurements are indicated by . . . . Only representative Fe I and FeH lines of 39 total shown. This table is available in machine-readable form.}
\end{minipage}

\end{table*}

\subsection{Uncertainties}
\label{sec:interferometric}

The uncertainties in \T\ and log $g$ were estimated by propagating the uncertainties in the iron abundances. In the right panel of Figure~\ref{fig:specteff}, the gray lines represent the variation in the Fe I and FeH abundance trends when their respective uncertainties are taken into account. The intersection between these bands defines a region of acceptable solutions, from which we estimate the uncertainty in \T~ (Figure~\ref{fig:specteff}, top panel). The resulting range in effective temperature corresponds to values within this region of typical uncertainty of $\sim$80 K. An analogous procedure is applied to determine the uncertainty in log $g$, leading to a typical uncertainty of $\sim$0.20 dex.

For the uncertainties in the derived iron abundances, we adopted the values reported by \citet{Melo2024} for M dwarfs and \citet{Souto2018} for G dwarfs. 
The typical uncertainties in iron abundances derived from Fe I lines are $\sim$0.09--0.13 dex, while those derived from FeH lines are systematically lower ($\sim$0.04--0.08 dex). 

\section{Discussion}\label{sec:baseline}

\subsection{Results with the Baseline Line List}

\begin{figure*}[!t]
    \centering
    \includegraphics[width=0.32\textwidth]{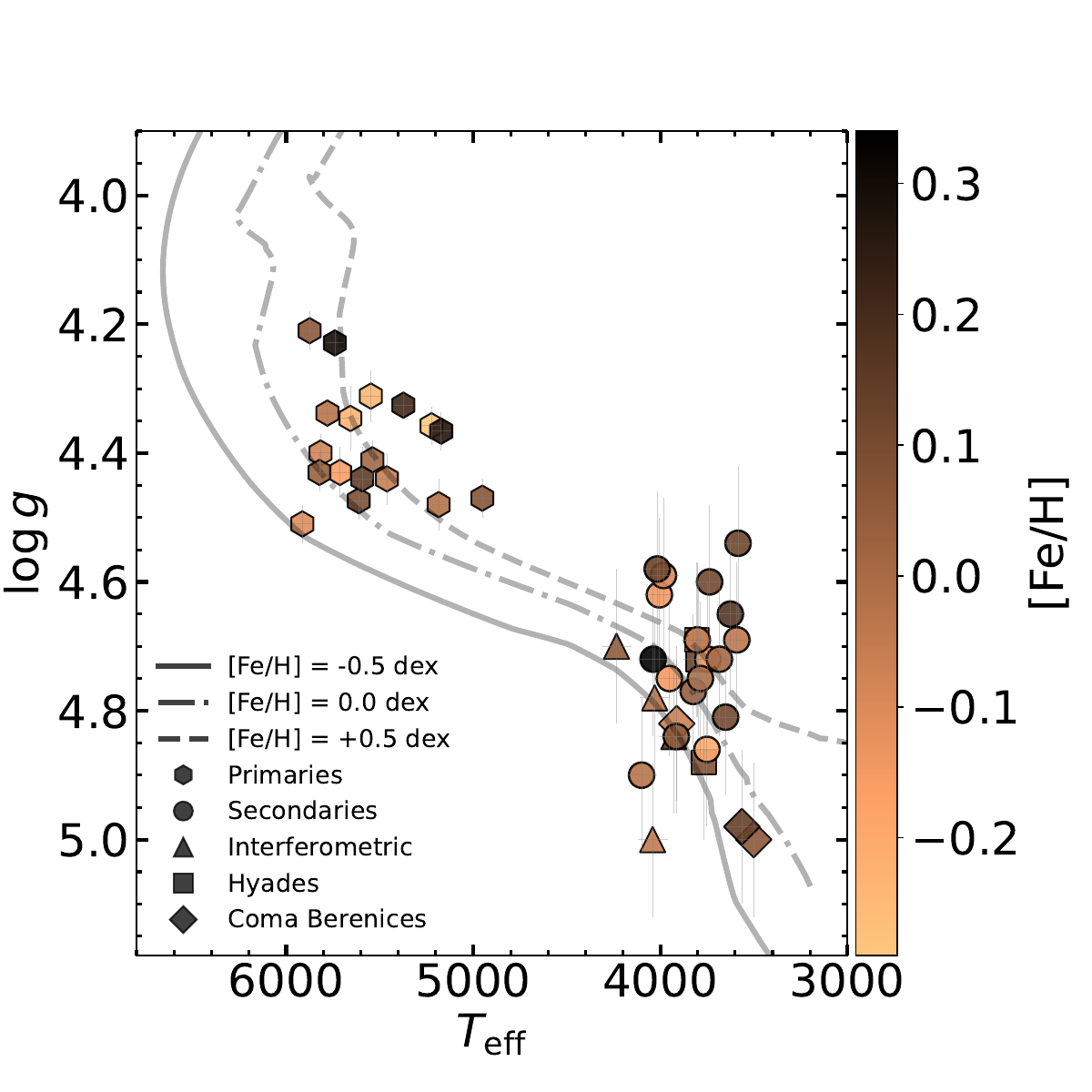}
    \includegraphics[width=0.32\textwidth]{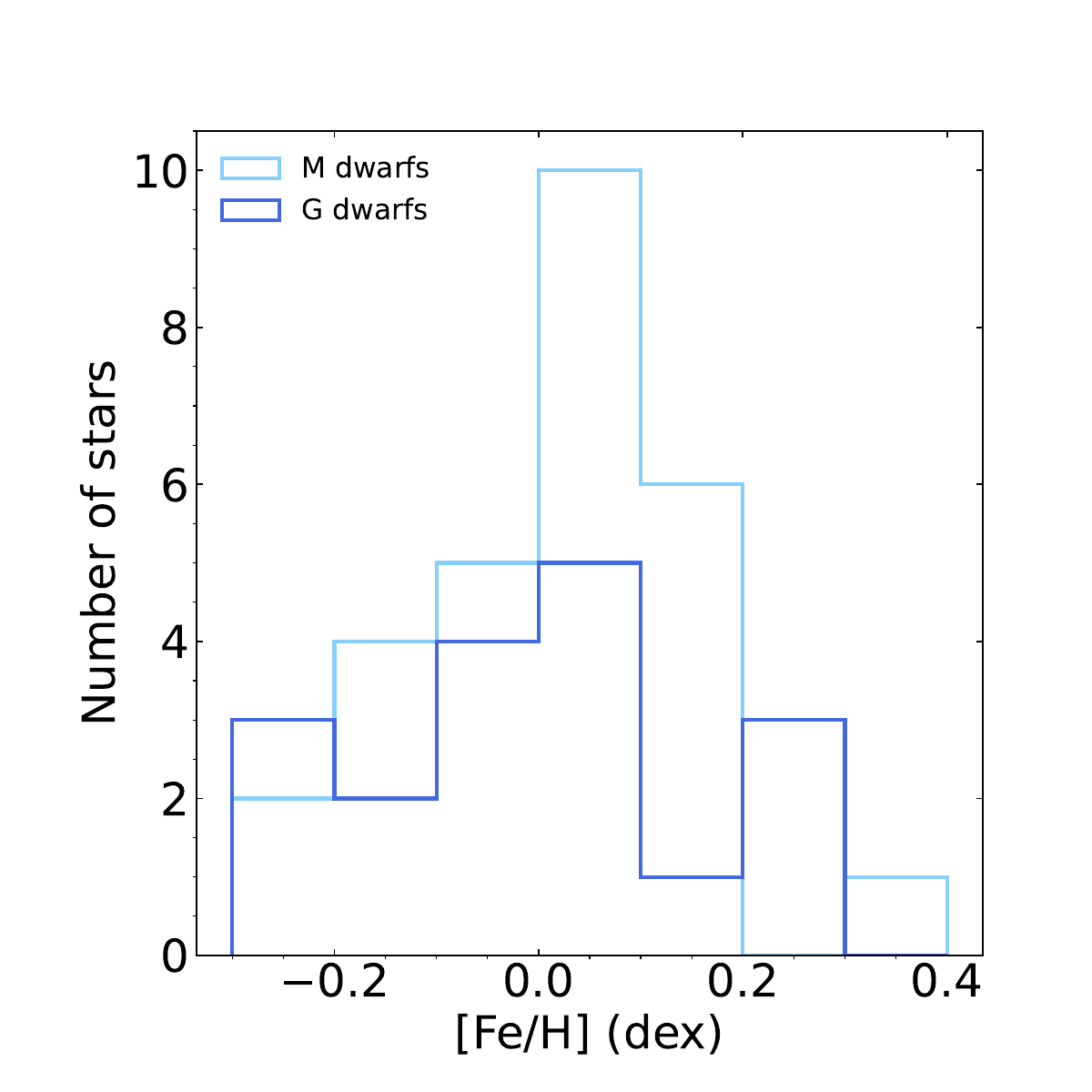}
    \includegraphics[width=0.32\textwidth]{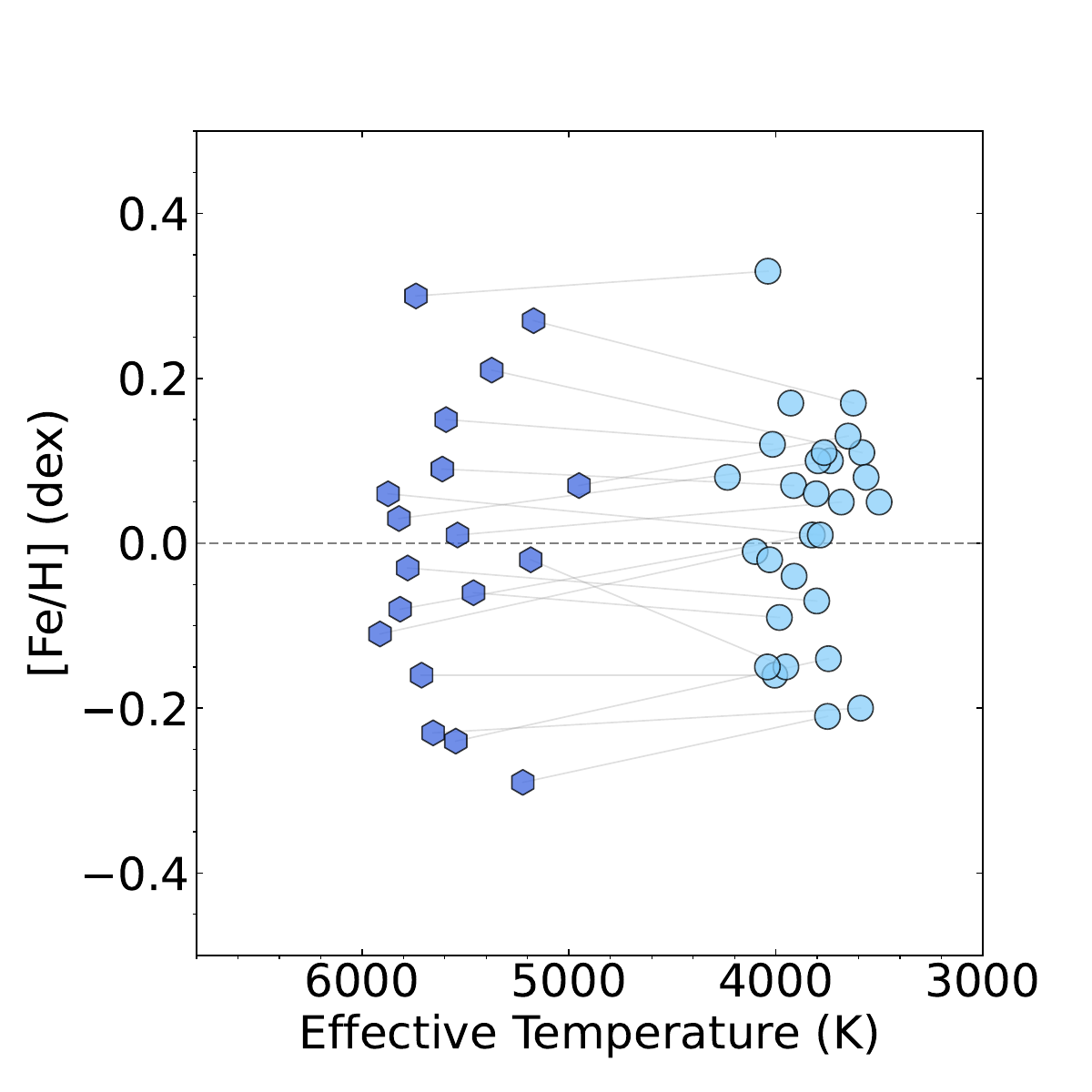}\\    
    \includegraphics[width=0.32\textwidth]{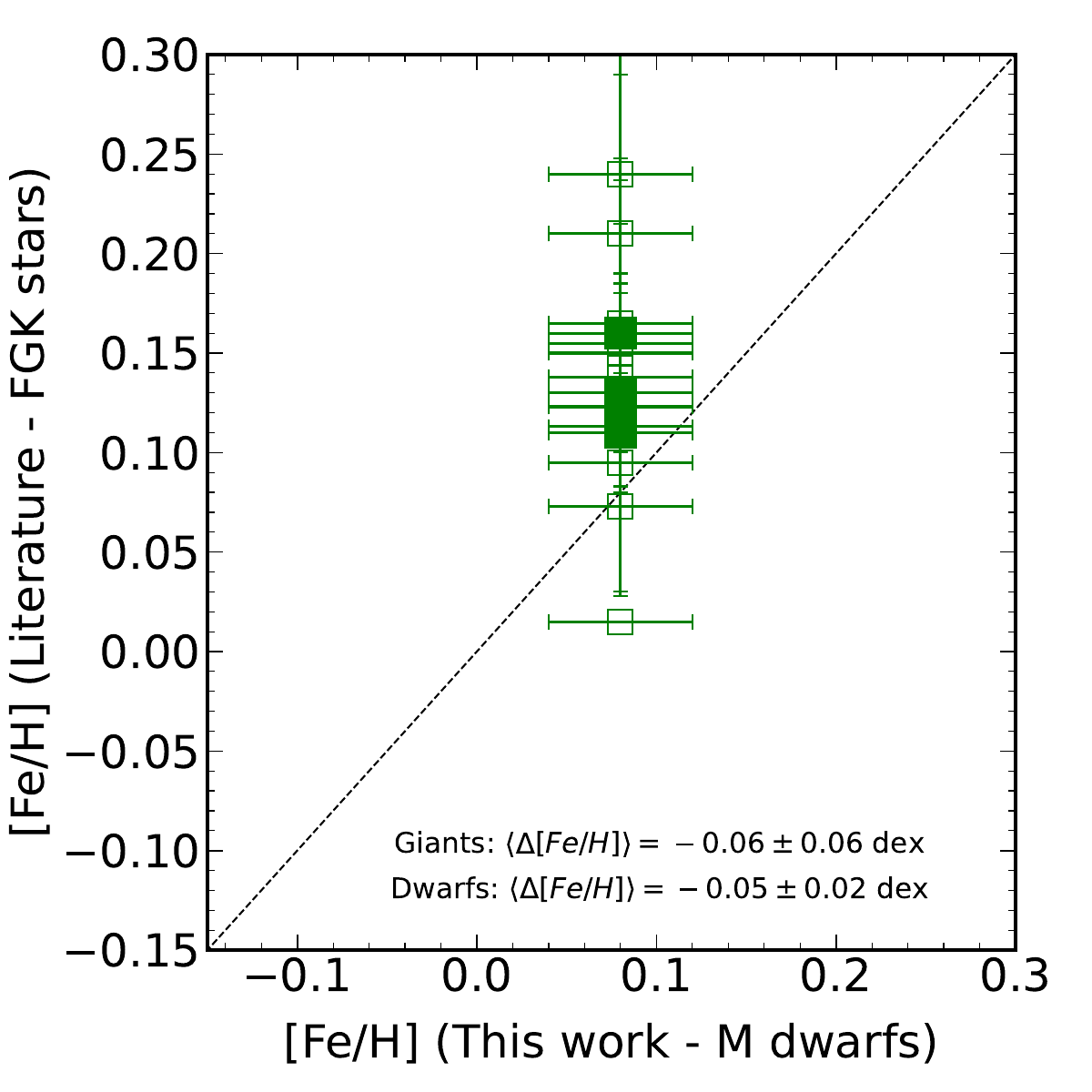}
    \includegraphics[width=0.32\textwidth]{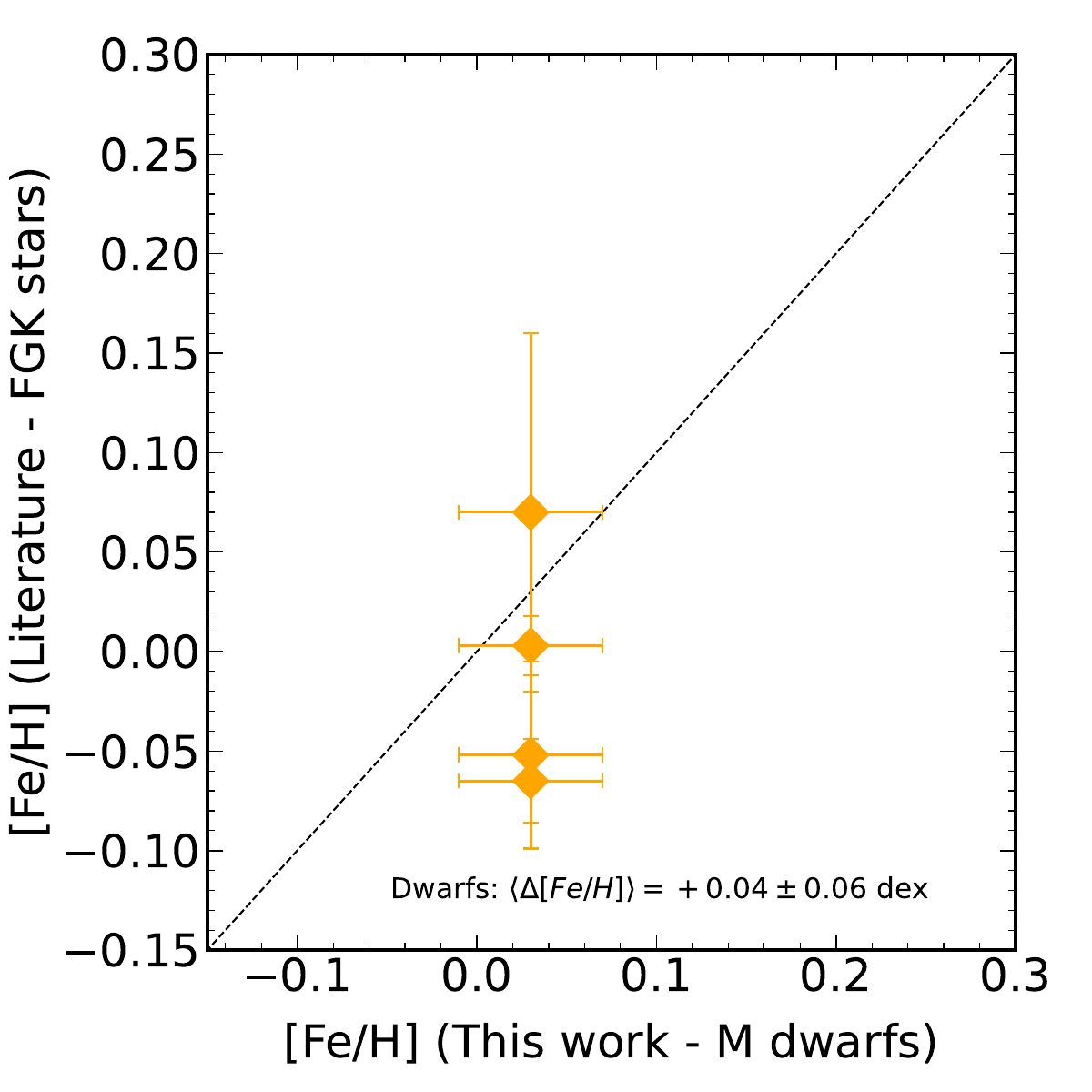}
    \caption{Kiel diagram displaying the stellar parameters of the M dwarfs obtained with the baseline APOGEE line list, along with adopted parameters from ASPCAP for the G dwarfs studied. Three MIST isochrones are shown for metallicities of --0.5, 0.0, and +0.5 dex, solar age (4.5~Gyr). Stellar parameters for dwarf stars, despite scattering due to uncertainties, are in general agreement with theoretical isochrones.
    {\it Top-middle:} Histogram of the metallicities derived for M (light blue) and G (dark blue) dwarf stars.  Both [Fe/H] distributions are in general agreement.
    {\it Top-right:} The distribution of the derived [Fe/H] as a function of the effective temperature. The absence of a dependence of [Fe/H] on \T~argues in favor of the good quality of our determinations. In light gray, we show a solid line connecting the binary pairs.
    {\it Bottom-left:} the mean metallicity of the Hyades obtained for the M dwarfs in this study in comparison with optical results from the literature. {\it Bottom-right panel:} The same as bottom-left, but for the Coma Berenices open cluster.  
     }    
    \label{fig:hist_teff_feh}
\end{figure*}

The fundamental parameters of T$_{\rm eff}$, log $g$, and [Fe/H] derived for the M dwarfs in this study were obtained using the original baseline APOGEE line list (\citealt{Smith2021}, we note that the f-values of atomic lines were tuned to the solar and Arcturus spectra, whereas molecular lines were not). These effective temperatures and surface gravities are given in Table \ref{stellaratm} and are shown as a Kiel diagram in the top-left panel of Figure \ref{fig:hist_teff_feh}, which includes all stars from our sample, noting that the MWM APOGEE DR19 values of \T~and log~$g$ are plotted for the G dwarfs. 
Stellar metallicities are indicated by the color bar, and the isochrones shown are the same as those used in Figure \ref{fig:CMD}. 
The values of T$_{\rm eff}$ and log~$g$ for the G and M dwarfs generally follow the MIST isochrone tracks, although with some scatter, which is of the order of the estimated uncertainties.

The top-middle panel of Figure \ref{fig:hist_teff_feh} presents the M-dwarf metallicities as a histogram (light blue), revealing a peak in the metallicity distribution near, or slightly above solar, for this small sample of nearby M-dwarfs. The dark blue histogram shows the corresponding metallicity distribution for the G dwarfs in the 18 binary systems, which closely overlaps that of the M dwarfs.
An important assessment pertaining to the overall accuracy of the derived iron abundances is provided by [Fe/H] versus \T, shown in the top-right panel of Figure \ref{fig:hist_teff_feh}, where the abundances of the M and G dwarfs exhibit no significant trends with effective temperature.  
The lack of any discernible trend with T$_{\rm eff}$ suggests that the abundance results are robust, do not suffer from significant systematics, and that non-LTE effects are likely small. 

Focusing on the open cluster members included in the sample, the mean metallicities for the M dwarfs are: $\langle$[Fe/H]$\rangle$ = +0.08 $\pm$ 0.04 dex for the Hyades and $\langle$[Fe/H]$\rangle$ = +0.02 $\pm$ 0.08 dex for Coma Berenices. Both cluster metallicities are on the metal-rich side of the distribution. 
In the bottom-left panel of Figure \ref{fig:hist_teff_feh}, we show a comparison of the mean [Fe/H] obtained for the Hyades M dwarfs in our sample with results for warmer stars from different high-resolution optical studies in the literature. We selected studies having red giants (\citealt{FernandezVillacanas1990}, \citealt{McWilliam1990}, \citealt{Smith1999}, \citealt{LuckChallener1995}, \citealt{Mishenina2006}, \citealt{daSilva2006}, \citealt{Schuler2006}, \citealt{Hekker2007}, \citealt{Carrera2011}, \citealt{Kang2011}, \citealt{Tabernero2012}, \citealt{Mortier2013}, \citealt{daSilva2015}, \citealt{Maldonado2016}), as well as comparison literature results for warmer dwarfs (\T~between 5000 and 7000 K) from \cite{Paulson2003}, \cite{Takeda2013}, and \cite{Liu2016}. 
The mean metallicity for the dwarfs from the literature is $\langle$[M/H]$\rangle$ = +0.13 $\pm$ 0.04 dex, while the mean value for the giants is +0.14 $\pm$ 0.07 dex. In principle, the iron abundances of red giants should be minimally affected by atomic diffusion and provide better comparisons for the metallicity scale of the M dwarfs, although in practice, the respective mean literature results for the Hyades are indistinguishable, indicating agreement within the uncertainties.  
The corresponding value for our three Hyades M dwarfs is $\langle$[Fe/H]$\rangle$ = +0.08 $\pm$ 0.04 dex. The differences between our value of $\langle$[Fe/H]$\rangle$ from the Hyades M dwarfs, derived using the baseline FeH line list, and those of the warmer giant and dwarf Hyades members of $\langle$$\Delta$[Fe/H]$\rangle$(This study - Literature) = -0.06 and -0.05 dex are within the uncertainties of all of the various studies.
For Coma Berenices (bottom-right panel of Figure \ref{fig:hist_teff_feh}), the case is similar, with our average result for the M dwarfs also comparing well, within the uncertainties, with literature metallicities for warmer Coma Berenices members taken from \cite{Cayrel1990}, \cite{Friel_Boesgaard1992}, \cite{Gebran2007}, and \cite{Aguero2025}. 

\subsection{Validating the FeH line list using benchmark M Dwarf Stars}\label{sec:inter_stars}

One goal of this study is to test the accuracy of the FeH E-A$^{4}\Pi$ line $gf$-values adopted in the APOGEE line list (\citealt{Smith2021}), as estimates of $gf$-value accuracies for this line system in FeH are difficult to ascertain (e.g., \citealt{Hargreaves2010}).  
To do so, we investigated trends in stellar parameters and Fe-abundances derived using FeH lines using three different tests involving the three different types of benchmarks in this study: the interferometric sample of four stars, the 18 M dwarfs in binary systems with G-dwarf primaries, and the M-dwarf members of Coma Berenices and the Hyades. 

\subsubsection{Stars with interferometric angular diameters}\label{sec:inter stars} 

\begin{figure}
    \centering
    \includegraphics[width=0.47\textwidth]{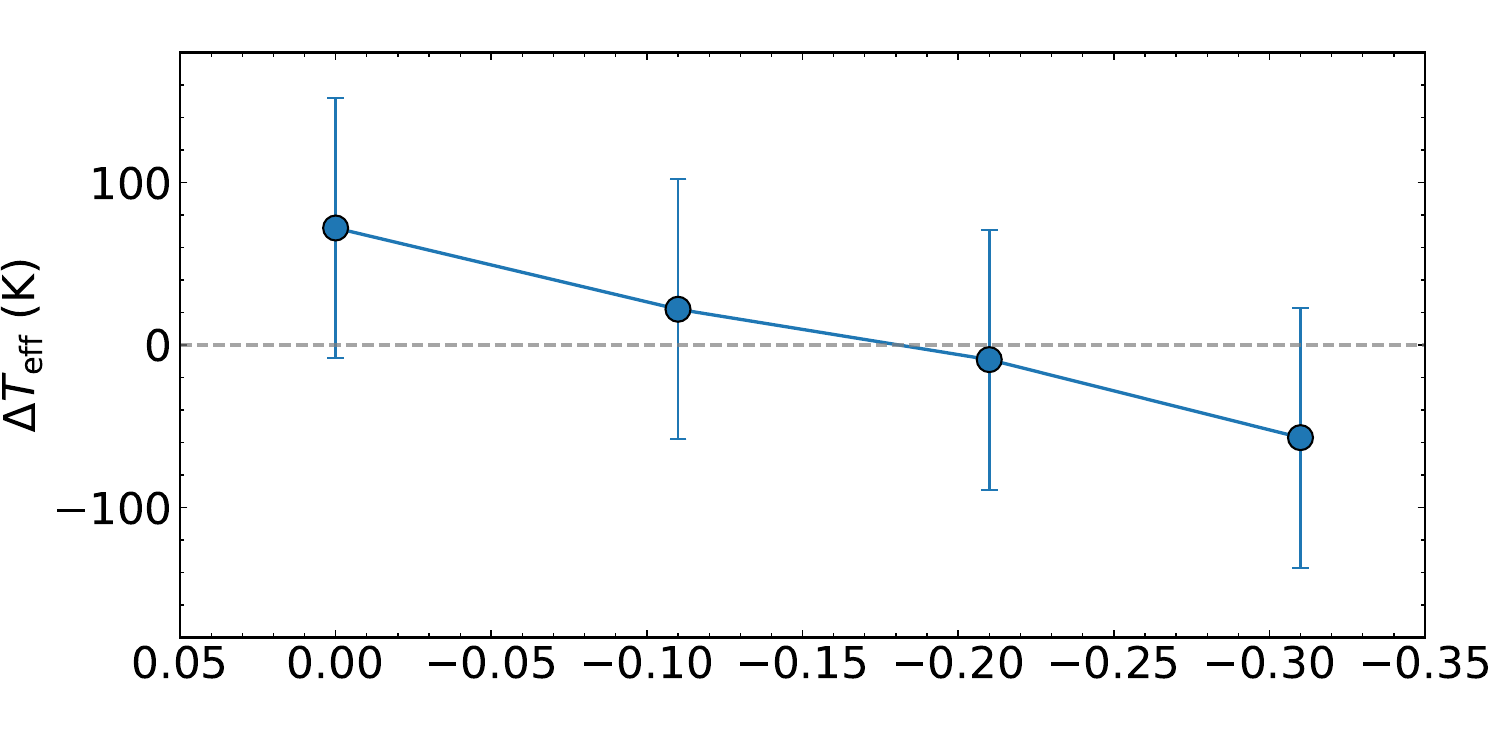}\vspace{-0.3cm}
    \includegraphics[width=0.47\textwidth]{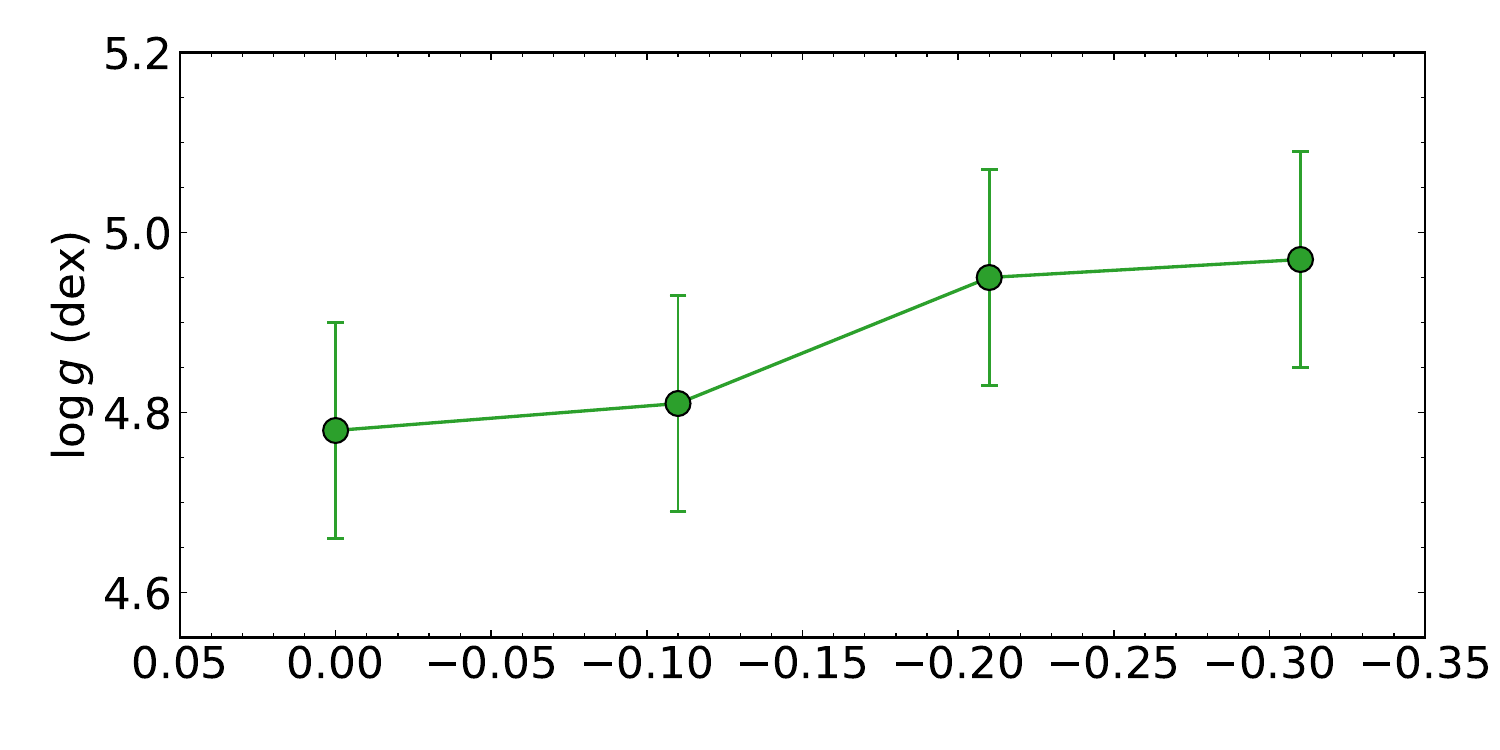}
    \includegraphics[width=0.47\textwidth]{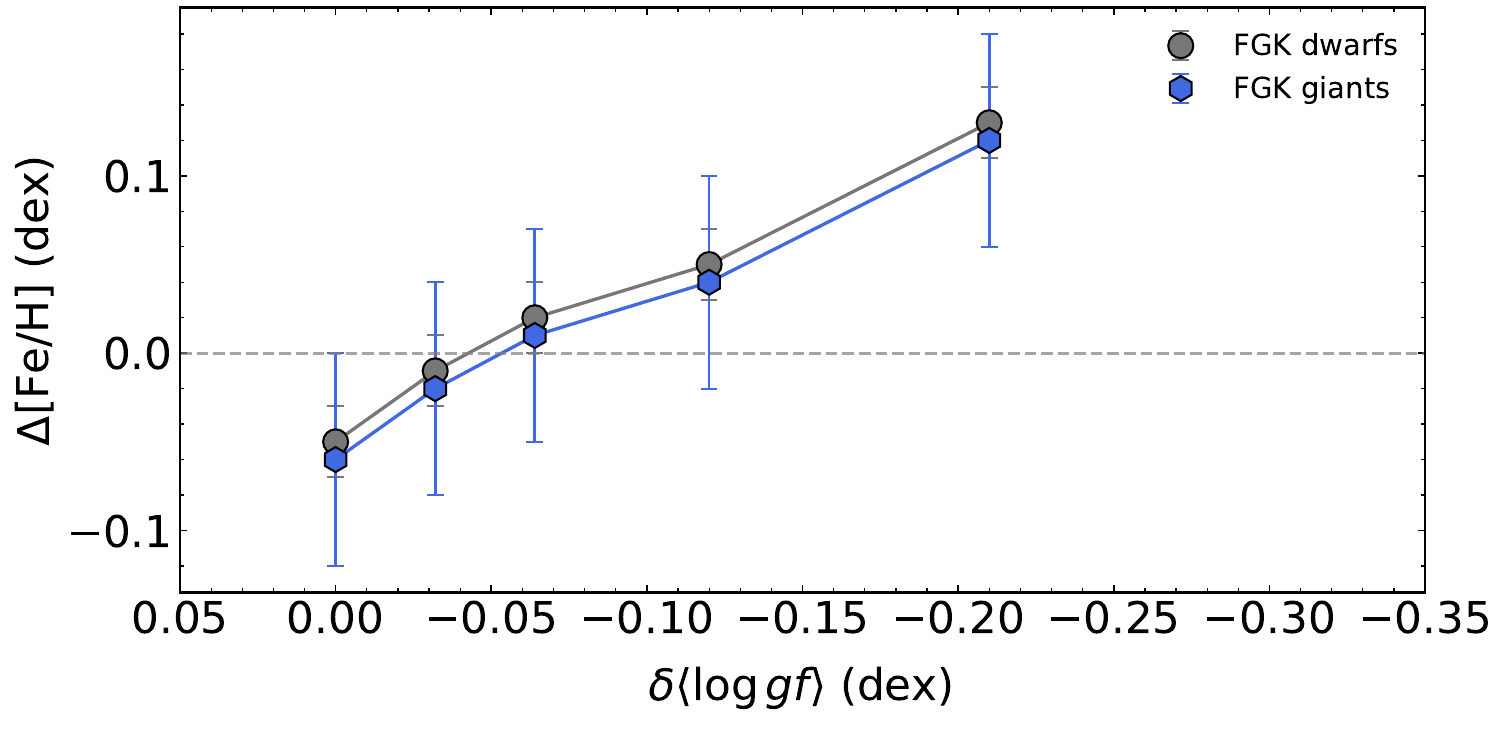}
    \caption{{\it Top panel}: The difference between the mean effective temperature from iron lines and the mean effective temperature calculated from angular diameters for benchmark M dwarfs as a function of systematic changes applied to the log $gf$ of the FeH line list. {\it Middle panel}: the derived log $g$ values for each realization in delta log $gf$. {\it Bottom Panel}: the difference between the average metallicity obtained in this study for the Hyades M dwarfs and the average metallicity of literature values for Hyades giants and warmer dwarfs as a function of $\delta$log $gf$.}
    \label{fig:deltas_gf} 
\end{figure}

Four of the benchmark M dwarfs (Table \ref{stellaratm}) have measured interferometric angular diameters in the literature. Using these interferometric angular diameter measurements from \cite{Boyajian2012}, along with the latest parallax data from Gaia DR3 (\citealp{Gaia_eDR3}), and luminosities from \cite{Mann2013}, we derived stellar radii and effective temperatures for these four stars.  The values of T$_{\rm eff}$ from interferometrically-derived stellar disks are employed as fiducial points with which to investigate our temperature scale derived from the diagnostic combination of Fe I and FeH lines.    
Comparing the effective temperatures derived spectroscopically from the \T-A(Fe) technique employed here with those determined from interferometric data, we find an offset of $\delta$(\T)$_{\rm[This~work - Benchmark]}$ = +72 $\pm$ 79 K, which is consistent within the estimated uncertainties.  This suggests that uncertainties in the accuracy of the FeH $g$f-values used here are not extremely large, but that estimates of the changes required to bring the two effective temperature scales closer to agreement are useful. 

To quantify the dependence of the derived T$_{\rm eff}$ on the adopted $g$f-values of the FeH transitions, we applied average adjustments, lowering by the same amount the log $gf$-values of the E-A$^{4}\Pi$ FeH transitions across the line list.
We note that while the $gf$-values of the E-A$^{4}\Pi$ FeH lines were systematically adjusted, the $gf$ values of the Fe I transitions were kept unchanged, and this decision was based on the previous extensive validation of Fe I lines performed by the APOGEE team using the solar and Arcturus spectra as benchmarks, as discussed in \citet{Smith2021}. 

The top and middle panels of Figure \ref{fig:deltas_gf} show the sensitivity of the derived parameters \T~ and log $g$ to changes in the $gf$ values of the E-A$^{4}\Pi$ FeH lines. Lowering all log $gf$-values by 0.1 dex led to excellent agreement between the spectroscopic and interferometric \T~values, with the mean $\delta$(\T)$_{\rm [This~work - Interferometric]}$ = +22 $\pm$ 80 K.
An additional 0.1 dex reduction resulted in excellent agreement between the spectroscopic and interferometric T$_{\rm eff}$-scales: $\delta$(log $gf$) = -0.2 leads to $\delta$(\T)$_{\rm [This~work - Interferometric]}$ = -9 $\pm$ 81 K. An additional change of $\delta$(log $gf$) by -0.1 dex resulted in a difference of 50 K in the derived mean effective temperatures (from 4058 K with the original line list to 4008 K with the changed line list.)

Given the spectroscopic methodology adopted in this study to obtain stellar parameters, adjustments to the $gf$ values of FeH E-A$^{4}\Pi$ lines also affect the derived surface gravity values. The middle panel of Figure \ref{fig:deltas_gf} shows the sensitivity of the log $g$ determinations to $\delta$(log $gf$). Unlike the effective temperature, which becomes smaller, an overall decrease in the log $gf$ values causes an increase in the derived log $ gs$, as shown in the middle panel of Figure \ref{fig:deltas_gf}. 
There is a very small change in log $g$ for a $\delta$(log $gf$) = -0.1 dex (from $\langle$log $g$$\rangle$ = 4.78 dex with the original line list to $\langle$log $g$$\rangle$ = 4.81 dex with the modified line list); the difference in derived log $g$ is larger for $\delta$(log $gf$) = -0.2 dex, resulting in a log $g$ change of 0.18 dex, which is still within the uncertanties in our log $g$ determinations (Section \ref{sec:interferometric}). 
The bottom panel of Figure \ref{fig:deltas_gf} shows the same x-axis, but now we compare how the metallicity results from the Hyades open clusters change compared to the literature. We find that progressively larger modifications to the FeH log $gf$ values systematically increase the metallicity discrepancies with respect to the literature abundance scale. This behavior indicates that introducing significant modifications to the FeH line list produces artificial systematic offsets in the derived metallicities, suggesting that the \cite{Hargreaves2010} FeH log $gf$ values already provide a consistent metallicity scale for APOGEE spectra.

\subsubsection{Binary stars}\label{sec:binary stars}

\begin{figure*}[!t]
    \centering
    \includegraphics[width=0.32\textwidth]{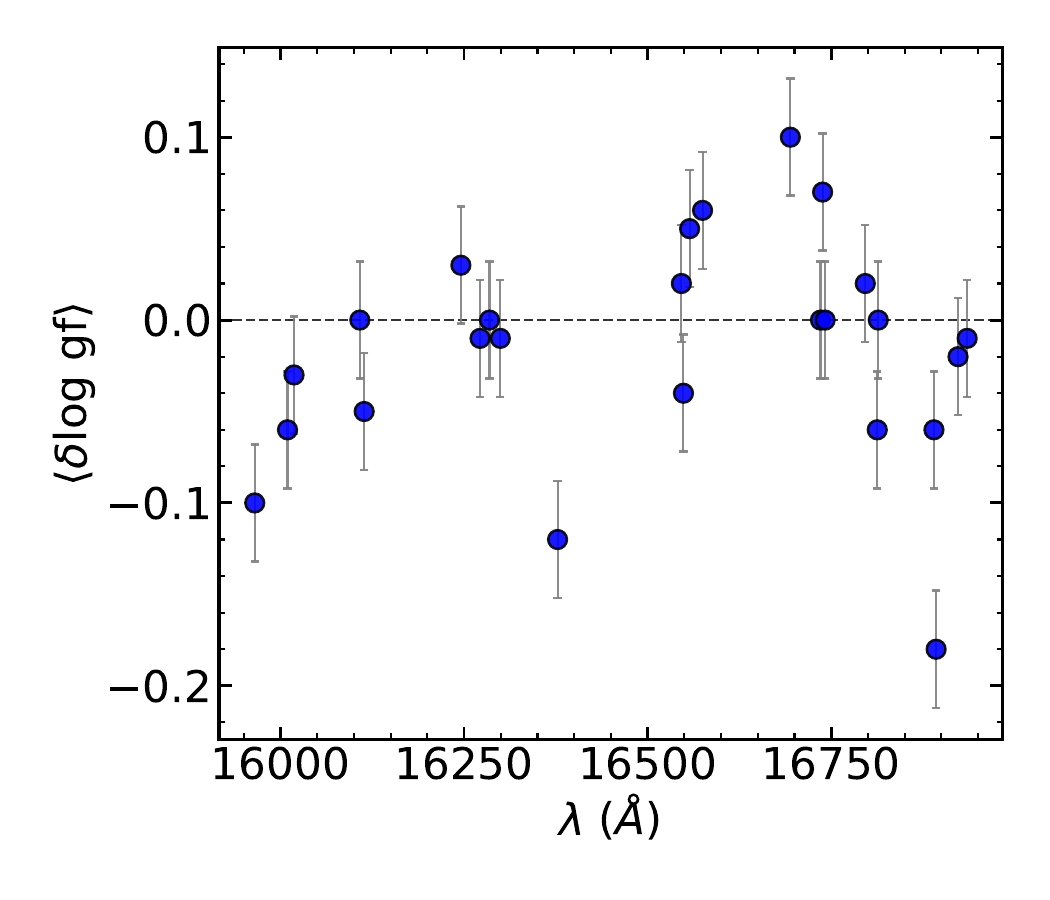}
    \includegraphics[width=0.32\textwidth]{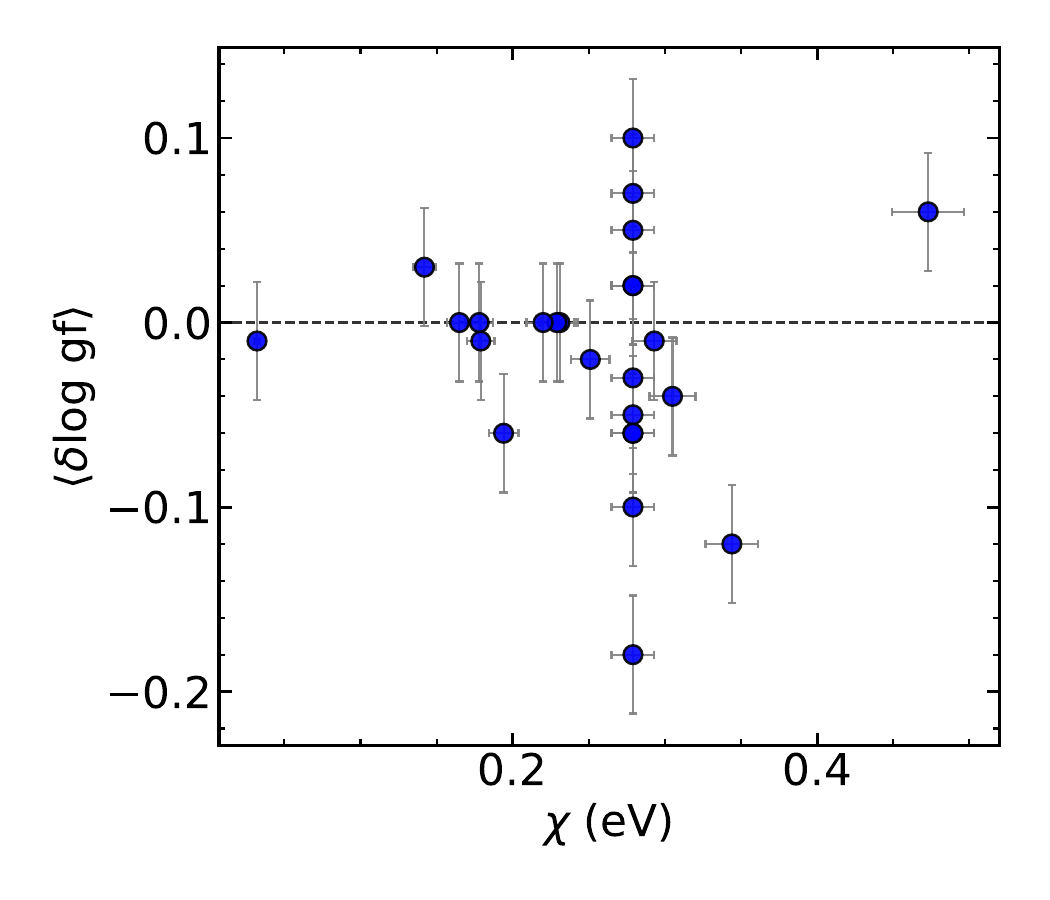}
    \includegraphics[width=0.32\textwidth]{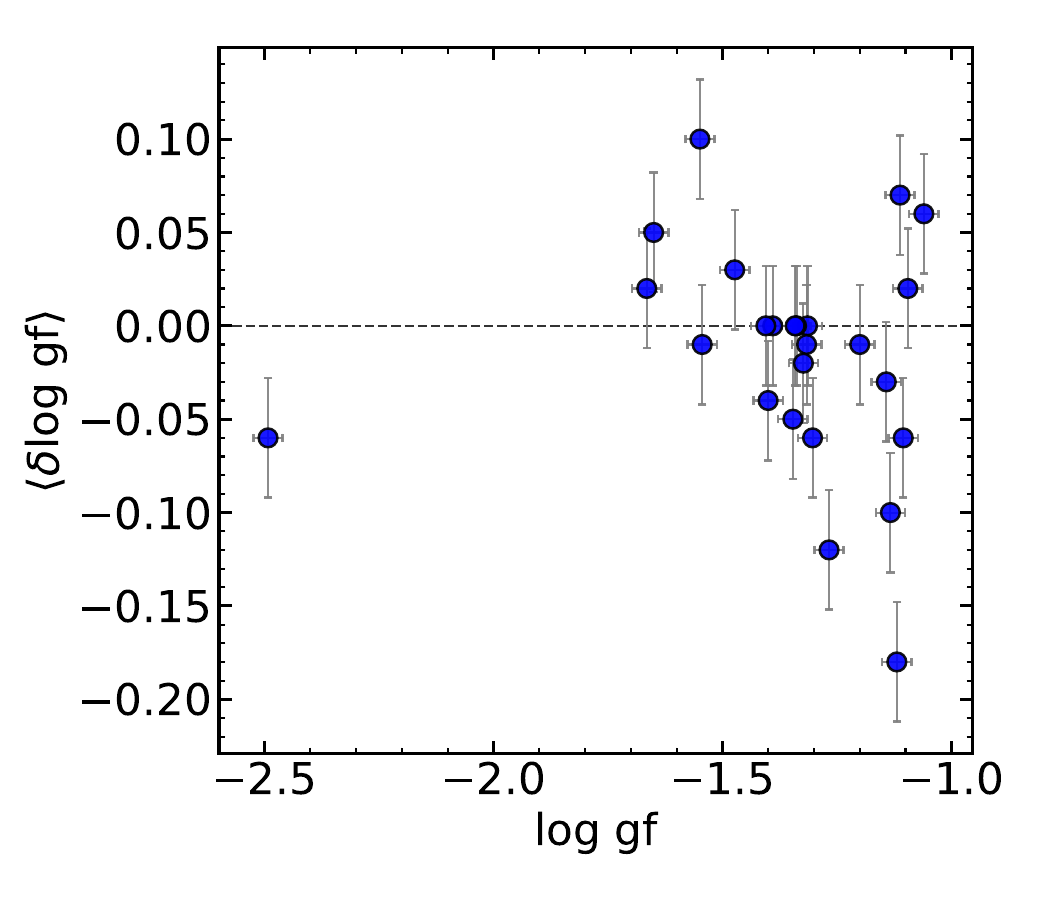}
    \caption{Line-by-line $\delta$(log $gf$) obtained for the binary stars for FeH transitions as a function of wavelength ($\lambda$), excitation potential ($\chi$), and log $gf$. The dashed horizontal line indicates no change in the FeH linelist. No significant trends are observed with any of the examined parameters.}
    \label{fig:loggf_lines} 
\end{figure*}

We performed a complementary analysis using the 18 binary systems in our sample to further gauge the FeH $gf$ values. In this analysis, iron abundances derived from Fe I lines in the G-dwarf primaries were adopted as benchmarks for metallicity.
We adjusted the log $gf$ values of the diagnostic FeH lines, which are all from the E-A$^{4}\Pi$ transition (Section \ref{sec:FeH}), until the iron abundances derived from FeH lines in the M-dwarf secondaries matched the A(Fe I) of their G-dwarf primaries. We note that separate log $gf$ adjustments were applied individually to each diagnostic FeH line in order to achieve convergence between the M- and G-dwarf abundances. This methodology is similar to one that derives astrophysical $gf$-values for individual transitions using benchmark stars, such as the Sun or Arcturus, as references for the abundance scales. The individual log $gf$ adjustments obtained for the diagnostic FeH lines in this study are in Table \ref{tab:delta_loggf}. 

\begin{table*}
\centering
\scriptsize
\setlength{\tabcolsep}{4pt}
\caption{Individual $\log gf$ adjustments for the diagnostic FeH lines}
\label{tab:delta_loggf}

\begin{minipage}{\textwidth}
\centering

\begin{tabular}{lccccccccccccc}
\hline\hline
2MASS~ID &
15965 & 16009.6 & 16018.5 & 16108.1 & 16114 & 16245.7 &
16271.8 & 16284.7 & 16299.4 & 16377.4 &
$\cdots$ &
$\langle \delta \log gf \rangle$ (dex) \\
\hline
\multicolumn{13}{l}{\textit{Secondaries}} \\
\hline
2M12414006+4103080 & \ldots & \ldots & \ldots &  0.00  & \ldots  & \ldots  &  0.00   & \ldots  & \ldots   & \ldots  & $\cdots$ & \ldots \\
2M06312373+0036445 & \ldots & \ldots & \ldots & \ldots  & \ldots  & \ldots  & \ldots  &  0.00   & $-$0.13 & \ldots  & $\cdots$ & \ldots \\
2M08083496+3047575 & \ldots & \ldots & \ldots &  0.00  &  0.03   & \ldots  &  0.00   & \ldots  & \ldots   & $-$0.04 & $\cdots$ & \ldots \\
2M08092559+5202190 & \ldots & \ldots & $-$0.07 & \ldots & \ldots & \ldots  & \ldots  & \ldots  & \ldots   & \ldots  & $\cdots$ & \ldots \\
2M08485678+1124111 & \ldots & \ldots & \ldots & \ldots  & \ldots  & \ldots  & $-$0.05 & \ldots  & \ldots   & \ldots  & $\cdots$ & \ldots \\
$\vdots$ & $\vdots$ & $\vdots$ & $\vdots$ & $\vdots$ & $\vdots$ & $\vdots$ & $\vdots$ & $\vdots$ & $\vdots$ & $\vdots$ & $\vdots$ & $\vdots$ \\
2M03044335+6144097 &  0.00  & $-$0.06 & \ldots & \ldots & \ldots & \ldots  &  0.05   & \ldots  & \ldots   & \ldots  & $\cdots$ & \ldots \\
\hline
\multicolumn{13}{l}{\textit{Summary statistics}} \\
\hline
Mean       & $-$0.10 & $-$0.06 & $-$0.03 & 0.00 & $-$0.05 & 0.03 & $-$0.01 & 0.00 & $-$0.01 & $-$0.12 & $\cdots$ & $-$0.016 \\
Std.\ dev. &   0.14  &   0.01  &   0.07  & 0.06 &   0.13  & 0.04 &   0.07  & 0.00 &   0.24  &   0.12  & $\cdots$ &   0.055  \\
\hline
\multicolumn{13}{l}{\textit{Coma Berenices}} \\
\hline
2M12241121+2653166 & $-$0.27 & \ldots & $-$0.15 & $-$0.08 & \ldots & $-$0.12 & $-$0.17 & $-$0.20 & \ldots & $-$0.27 & $\cdots$ & $-0.12 \pm 0.11$ \\
\hline
\end{tabular}

\vspace{2pt}
{\footnotesize \textbf{Note}. The full table is available in machine-readable form.}

\end{minipage}
\end{table*}

Figure \ref{fig:loggf_lines} shows the average difference in log $gf$ for each diagnostic FeH line as a function of the wavelength of the FeH line (left panel), the excitation potential, $\chi$, of the transition (middle panel), and the initial log $gf$ value from the original baseline line list. The average adjustment for all lines is $\langle\delta$log $gf\rangle$ = -0.016 $\pm$ 0.055 dex, reflecting general agreement between the metallicities of the G-type primaries and the M dwarfs. In addition, there are no significant trends in any of these diagrams, indicating internal consistency in the line list and suggesting that the astrophysical corrections to the log $gf$ are not driven by systematic dependencies on the individual line properties. This absence of correlations supports the assumption that the FeH transitions are well-characterized across the T$_{\rm eff}$-log $g$-[Fe/H] parameter space spanned by the M-dwarf secondaries and that any residual offsets are likely due to small, global zero-point differences rather than deficiencies in specific subsets of lines. 
  
We point out, however, that although the metallicities for the G-type primaries are, on average, indistinguishable from those of the M-dwarf secondaries, there remains the possibility that the metallicities of the G-dwarfs do not reflect those of their natal clouds due to the effects of atomic diffusion. 
The expectation from stellar models of atomic diffusion is that photospheric iron abundances (along with other ``metals'') decrease slightly in G-type stars as functions of both stellar mass and increasing age (e.g., \citealt{Dotter2017}). 
The downward diffusion of helium, along with the heavier metals, out of the outer convection zones of solar-type stars over time results in decreases in the surface mass fractions Y and Z, with an increase in X.  Models from \cite{Dotter2017} find $\Delta$[Fe/H]$\sim$-0.03 to -0.10 dex for G-type dwarfs (with the magnitude of the change depending on age, mass, and metallicity). Conversely, diffusion does not take place in the atmospheres of M~dwarfs, as they have fully convective envelopes \citep[e.g.,][]{baraffe2018}, so that their chemical compositions probably reflect those of the clouds they were formed from.

\begin{figure}[!t]
\centering
     \includegraphics[width=0.45\textwidth]{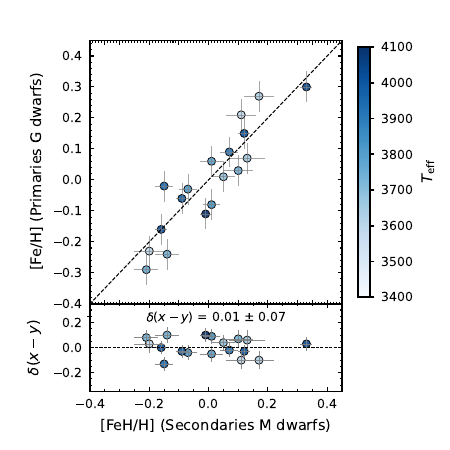}
     \caption{Comparison of [Fe/H] derived for G-dwarf primaries and their corresponding M-dwarf secondaries. The color bar represents the M dwarfs \T, and includes a residual diagram at the bottom.}
     \label{fig:abund}
\end{figure}

In Figure~\ref{fig:abund} we compare the metallicities derived for the secondary M dwarfs using FeH lines with those obtained for the primary G dwarfs from Fe I lines in the same binary systems. The excellent agreement between the two measurements demonstrates that the atmospheric parameters derived from the Fe I/FeH equilibrium method provide a consistent abundance scale across M and G dwarfs. We find a mean metallicity difference of only $\langle$$\delta(x-y)$$\rangle$$~= 0.01 \pm 0.07$ dex, with no significant systematic trend over the metallicity range explored or in effective temperature, as shown in the color bar of the Figure. This result indicates that iron abundances derived from FeH features in APOGEE M-dwarf spectra are fully compatible with those obtained from atomic Fe I lines in warmer stars. A residual diagram is shown in the bottom panel of the Figure.

Finally, we note that \citet{Hargreaves2010} investigated the sensitivity of FeH line intensities to uncertainties in the lower-state energies of unclassified transitions. They showed that a $\pm$5\% variation in the lower energy level ($\pm112.5$ cm$^{-1}$) produces changes of less than $\pm$0.5\% in the calculated line intensities over a temperature variation of $\pm100$ K. Motivated by this analysis, we performed an additional sensitivity test by varying, on a line-by-line basis, the lower energy levels of the FeH transitions by $\pm5\%$, generating two modified line lists with recalculated log $gf$ values. We find that these perturbations result in a mean variation of only $\delta = \pm0.032$ dex in the derived log $gf$ values relative to the original FeH line list.

\subsubsection{Open cluster stars}\label{sec:open clusters} 

Open clusters provide excellent testbeds for stellar astrophysics, as one of their properties reflects the assumption that the metallicities of open cluster members form with the same chemical composition as that of their natal molecular cloud. 
In this section, we describe an analysis similar to that for the binaries (Section \ref{sec:binary stars}), but leveraging open cluster members as [Fe/H] benchmarks.  We first analyzed the FeH lines in one M-dwarf member of the Coma Berenices open cluster using the abundance results for the K-type dwarfs in Coma Berenices as metallicity benchmarks, as well as using literature results for warmer stellar members of the Hyades open cluster as metallicity benchmarks for three Hyades M-dwarfs.

In the first analysis, we selected the Coma Berenices M-dwarf member, 2M12241121+2653166, and adjusted the log $gf$-values of the diagnostic FeH transitions line-by-line to match the mean iron abundance for the cluster K-dwarfs of [Fe/H]$\sim$+0.04 $\pm$ 0.01 dex reported in \cite{Souto2021}, which was obtained from measurements of diagnostic Fe I lines in the APOGEE spectra of four K-dwarf stars. 
The individual log $gf$ adjustments of FeH lines required in this case are presented in Table~\ref{tab:delta_loggf}. We found that an average adjustment of $\delta$ (log $gf$) = -0.120 $\pm$ 0.11 dex was needed to place the FeH derived M dwarf iron abundances on the same scale as those obtained from Fe I lines in K dwarfs from \cite{Souto2021}. 

To further investigate whether astrophysical $gf$-value adjustments are needed for the APOGEE FeH line list, we compared independent iron abundance measurements from the literature for Hyades members. The Hyades open cluster has been well-studied and has numerous independent metallicity measurements in the literature based on high-resolution optical spectroscopy. A summary plot of the literature metallicities as a function of the effective temperature for the Hyades members is presented in Figure 4 of \cite{Wanderley2023} and shows a decrease in the metallicity for dwarf stars with T$_{\rm eff}\sim$6000--7000 K, which, in addition to possible systematic differences in abundance scales in the different studies, can be explained generally by atomic diffusion processes. 

Figure \ref{fig:deltas_gf} (bottom panel) illustrates the sensitivity of the FeH metallicities for the Hyades M dwarfs as a function of $\delta$(log $gf$), which, as previously discussed, represents an overall adjustment in the $gf$-values of the E-A$^{4}\Pi$ transitions derived using \citet{Hargreaves2010}.  We show the difference between our mean metallicity for the Hyades minus the mean metallicity obtained from the literature studies for giants (blue curve) and dwarfs (gray curve). As previously discussed, the average metallicities for the giants and warmer dwarfs in the selected papers are very similar, to within 0.01 dex. 
The formal best agreement between the metallicities corresponds to a small $\sim$-0.04 dex adjustment to the FeH log $gf$ values, although all results are generally consistent within the uncertainties, except for the $\delta$(log $gf$) = -0.21 dex case that shows a larger deviation.

\section{Conclusions}\label{sec:conclusions}

In this work, spectroscopic data from APOGEE/DR19 were utilized to investigate spectral-line parameters, primarily $gf$-values, for FeH lines that are used in spectroscopic analyses to derive effective temperatures (\T), surface gravities (log $g$), and iron abundances for M dwarfs. 
The FeH lines in question are associated with the E$^{4}\Pi$ - A$^{4}\Pi$ electronic transition presented and discussed by \cite{Hargreaves2010}. Stellar parameters were derived using a homogeneous application of the Fe I - FeH quantitative spectroscopic analysis technique (\citealp{Souto2017}) in three samples of 28 benchmark M dwarfs: four nearby examples with measured interferometric diameters, 18 M-dwarf secondary members of binary systems all having G-dwarf primaries, as well as six M-dwarfs that are members of two open clusters (three in the Hyades and three in Coma Berenices). The interferometric sample serves as useful effective temperature benchmarks due to the fundamental definition of \T, while both the G-dwarf primaries and the open cluster chemical abundances provide fiducial benchmarks for stellar metallicity ([Fe/H]).   

Effective temperatures derived in the four interferometric M-dwarf benchmarks using the Fe I - FeH technique were found to have marginally larger values of \T~than those found by combining the published angular diameter measurements with Gaia parallaxes and published luminosities, with a mean difference and standard deviation of $\langle\Delta$T$_{\rm eff}\rangle$ = +72 $\pm$ 79 K.  Forcing this mean difference to go to zero in the four stars was found to be possible by decreasing the log $gf$ values of the FeH lines by -0.18 dex. 
Although forcing our spectroscopic \T~to the interferometric scale by changing the FeH $gf$-values brings the effective temperature scales into agreement, by design, it turns out that this does not yield metallicities that are consistent with those of the G-dwarf primaries in binary systems, or with the open-cluster benchmarks. In contrast, using the original \citet{Hargreaves2010} FeH line list without $gf$ value adjustments provides the most statistically robust agreement in the values of [Fe/H] for the binary and, to a lesser degree, the open cluster sample. 
Our analysis indicates that, given overall uncertainties in effective temperatures and metallicities, no modifications to the log $gf$ values reported by \citet{Hargreaves2010} are required and we therefore suggest adopting the original FeH line list based on the results presented in \citet{Hargreaves2010}.

\begin{acknowledgments}
We thank the referee for comments that helped improve the paper. A.S.A acknowledges support from the fellowship by Coordenação de Aperfeiçoamento de Pessoal de Nível Superior - CAPES.
D.S. acknowledges support from the Foundation for Research and Technological Innovation Support of the State of Sergipe (FAPITEC/SE) and the National Council for Scientific and Technological Development (CNPq), under grant numbers 794017/2013 and 444372/2024-5.
K.C. and V.S. acknowledge support from the NASA ADAP grant 80NSSC23K1410.
V.L.T. acknowledges support from the CNPq through the Postdoctoral Junior (PDJ) fellowship, process No. 152242/2024-4. C.A.P. acknowledges financial support from the Spanish Ministry of Science, Innovation and Universities (MICIU) projects PID2020-117493GB-I00, PID2023-149982NB-I00 and PID2023-146453NBI00. Funding for the Sloan Digital Sky Survey V has been provided by the Alfred P. Sloan Foundation, the Heising-Simons Foundation, the National Science Foundation, and the Participating Institutions. SDSS acknowledges support and resources from the Center for High-Performance Computing at the University of Utah. SDSS telescopes are located at Apache Point Observatory, funded by the Astrophysical Research Consortium and operated by New Mexico State University, and at Las Campanas Observatory, operated by the Carnegie Institution for Science. The SDSS website is \url{www.sdss.org}.
SDSS is managed by the Astrophysical Research Consortium for the Participating Institutions of the SDSS Collaboration, including Caltech, The Carnegie Institution for Science, Chilean National Time Allocation Committee (CNTAC) ratified researchers, The Flatiron Institute, the Gotham Participation Group, Harvard University, Heidelberg University, The Johns Hopkins University, L’Ecole polytechnique f\'{e}d\'{e}rale de Lausanne (EPFL), Leibniz-Institut f\"{u}r Astrophysik Potsdam (AIP), Max-Planck-Institut f\"{u}r Astronomie (MPIA Heidelberg), Max-Planck-Institut f\"{u}r Extraterrestrische Physik (MPE), Nanjing University, National Astronomical Observatories of China (NAOC), New Mexico State University, The Ohio State University, Pennsylvania State University, Smithsonian Astrophysical Observatory, Space Telescope Science Institute (STScI), the Stellar Astrophysics Participation Group, Universidad Nacional Aut\'{o}noma de M\'{e}xico, University of Arizona, University of Colorado Boulder, University of Illinois at Urbana-Champaign, University of Toronto, University of Utah, University of Virginia, Yale University, and Yunnan University.
\end{acknowledgments}

\vspace{5mm}
\facilities{Sloan.}

\software{BACCHUS \citep{Masseron2016} and TURBOSPECTRUM (\citealp{AlvarezandPlez1997}; \citealp{Plez2012}; \url{https://github.com/bertrandplez/Turbospectrum2019.git}).}

\bibliography{sample701}{}
\bibliographystyle{aasjournalv7}

\end{document}